\documentclass[5p]{elsarticle}

\graphicspath{{./figures/}}
\DeclareGraphicsExtensions{.pdf,.jpeg,.png,.jpg}
\usepackage{amssymb}
\usepackage{amsmath}
\usepackage{url}
\usepackage{aas_macros}
\usepackage[utf8]{inputenc}
\usepackage{color}

\journal{Astronomy and Computing}

\begin{document}

\begin{frontmatter}



\title{A New View of Observed Galaxies through 3D Modelling and Visualisation}


\author[l1]{Tim Dykes\footnotemark}
\author[l2]{Claudio Gheller}
\author[l3]{B\"arbel S. Koribalski}
\author[l4]{Klaus Dolag}
\author[l5]{Mel Krokos}

\address[l1]{HPE HPC/AI EMEA Research Lab, Broad Quay House, Broad Quay, Bristol, U.K.}
\address[l2]{Institute of Radioastronomy, INAF. Via P. Gobetti, 101 40129 Bologna, Italy}
\address[l3]{CSIRO Astronomy and Space Science, Australia Telescope National Facility, P.O. Box 76, Epping, NSW 2121, Australia.}
\address[l4]{Universit\"ats-Sternwarte M\"unchen, Scheinerstr.1, D-81679 M\"unchen, Germany.}
\address[l5]{School of Creative Technologies, University of Portsmouth, Portsmouth PO1 2DJ, U.K.}
\begin{abstract}
Observational astronomers survey the sky in great detail to gain a better understanding of many types of astronomical phenomena. In particular, the formation and evolution of galaxies, including our own, is a wide field of research. Three dimensional (spatial 3D) scientific visualisation is typically limited to simulated galaxies, due to the inherently two dimensional spatial resolution of Earth-based observations. However, with appropriate means of reconstruction, such visualisation can also be used to bring out the inherent 3D structure that exists in 2D observations of known galaxies, providing new views of these galaxies and visually illustrating the spatial relationships within galaxy groups that are not obvious in 2D. We present a novel approach to reconstruct and visualise 3D representations of nearby galaxies based on observational data using the scientific visualisation software Splotch. We apply our approach to a case study of the nearby barred spiral galaxy known as M83, presenting a new perspective of the M83 local group and highlighting the similarities between our reconstructed views of M83 and other known galaxies of similar inclinations.
\end{abstract}

\begin{keyword}
galaxy modelling \sep visualisation 

\end{keyword}

\end{frontmatter}


\section{Introduction}
\label{sec:introduction}

\begin{figure} [ht]
\begin{center}
\includegraphics[width=\columnwidth]{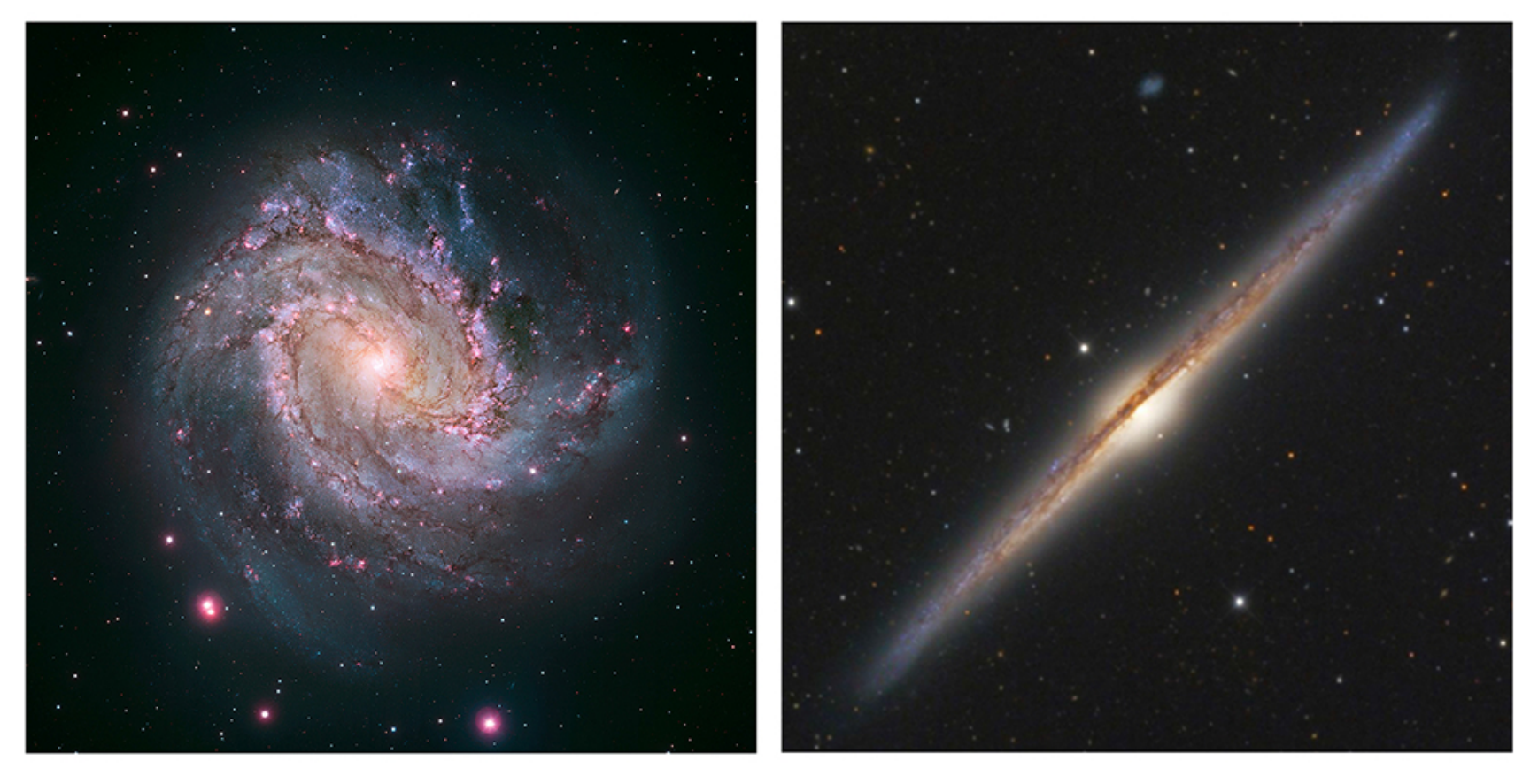}
\caption[Face-on v.s. edge-on galaxies.]{Optical images of two very similar spiral disk galaxies seen at different inclination angles. The {\bf face-on} view of M\,83 (\emph{left}) reveals the full complexity of its stellar spiral structure mixed with dust and young star-forming regions. In contrast, the {\bf edge-on} view of NGC~4565 (\emph{right}) reveals an old central bulge and strongly obscuring dust lanes seen against the thin stellar disk. --- Images courtesy of \emph{NASA, ESA, and the Hubble Heritage Team (STScI/AURA)} and \emph{Robert Franke}\footnotemark, respectively. --- North is up and East to the left.}
\label{fig:M83_NGC4565}
\end{center}
\end{figure}

\footnotetext{Corresponding author: tim.dykes@hpe.com}
The study of galaxy formation and evolution is a wide research field in astronomy, ranging from multi-frequency observations of the Milky Way and nearby galaxies to distant galaxy groups and clusters \cite{Meureretal06, GildePazetal07, Koribalski18}, complemented by cosmological galaxy simulations from the Big Bang to the present day \cite{Mcalpineetal16, Potteretal17}. While specific galaxies can only be observed from a single viewpoint on (or near) Earth, their intrinsic 3D shapes are well known from all-sky surveys, such as the Sloan Digital Sky Survey (SDSS) \cite{YorkEtAl00}, containing millions of objects at all possible orientation angles. As an example, Figure~\ref{fig:M83_NGC4565} highlights the contrasting views of two similar spiral disk galaxies seen at very different angles (represented by their inclination angle $i$): the barred galaxy M\,83 (\emph{left}) is seen nearly face-on ($i \sim 0$ degrees), while the galaxy NGC~4565 (\emph{right}) is seen nearly edge-on ($i \sim 90$ degrees). 

Detailed observations of galaxies over a large range of frequencies (e.g. optical, radio, infrared and ultraviolet) allow the measurement and derivation of many different properties, together enabling us to study galaxy morphology, kinematics, composition, mass, age, and formation history. Conversely, simulation of astronomical objects has been an active research topic in astronomy for many years \cite{peebles70, Vogelsberger20} with an aim to gain a better understanding of physical processes occurring over timescales so long, and across spatial scales so massive, that we cannot hope to observe them in real-time. In both cases, visualisation has been used extensively to help understand, analyse, and disseminate the results of such efforts, as discussed in Section \ref{sec:related}. In the context of the study of galaxies, visualisations are typically created from multi-dimensional data fields generated by numerical simulations and used to study properties of the source data; the focus is on presenting data in a manner facilitating intuitive comprehension rather than showing a visually realistic galaxy representation. Conversely, realistic 3D spatial visualisation of astronomical objects based on observations requires physically robust reconstruction methods paired with visualisation tools that support high quality rendering. 

This paper presents a novel methodology to reconstruct and visualise particle-based 3D models of nearby galaxies based on (a) observed multi-wavelength images and (b) detailed kinematic information. Our objectives are to: (1) create realistic views of galaxies including from viewpoints not otherwise possible to observe, (2) explore the validity of derived spatial 3D models of such galaxies, and (3) allow enhanced visual analysis of the 3D galaxy morphology (stars, gas and dust). The methodology is implemented within the open-source 3D scientific visualisation package Splotch\footnote{\url{https://wwwmpa.mpa-garching.mpg.de/~kdolag/Splotch/}} \cite{Dolagetal08}, enabling the reconstruction, visualisation, and exploration of individual, or groups of, recognisable galaxies. We expect that our pipeline may be utilised for scientific communication and outreach by generating immersive and realistic movies in 3D of known galaxies, allowing the non-expert viewer to grasp the connection between 2D observations and the real 3D structure of such objects. We also separate out components within that 3D structure, highlighting for example the large extent and warped morphology of the gaseous component that is often forgotten by both experts and non-experts alike. We hope that through physically realistic reconstruction and rendering this pipeline may also be used to compliment typical statistical analyses via objectives (2) and (3), allowing the astronomer to build better mental images of both the 3D structure of their object of study, and better comprehend spatial relationships between companion objects.

As a case study for our work, we reconstruct and visualise the nearby barred spiral galaxy known as M\,83, for its designation in the Messier catalogue, or as NGC~5236 in the New General Catalogue (NGC) of Nebulae and Clusters of Stars. As one of the closest and brightest spiral galaxies in the sky, M\,83 has a large collection of high resolution and multi-wavelength data publicly available, including H\,{\sc i} 21-cm data from the Local Volume H\,{\sc i} Survey (LVHIS; Koribalski et al. 2018). Furthermore, M\,83 is particularly well suited for visualisation due to its nearly face-on inclination and a large warped H\,{\sc i} disk, making it possible to use our sophisticated 3D reconstruction to visualise its likely edge-on appearance. The numerous dwarf companion galaxies surrounding M\,83 have also been modelled, allowing us to create a 3D visualisation of the galaxy group. 

The structure of this paper is as follows: Section \ref{sec:related} provides an overview of related work. Section \ref{sec:observations} presents a brief introduction to the techniques for galaxy observation and kinematic modelling that underpin our galaxy reconstruction approach. Section \ref{sec:modelling} introduces a new technique for reconstructing the physical structure of observed galaxies. Section \ref{sec:visualisation} describes our galaxy visualisation process, with results presented in Section \ref{sec:results}. Section \ref{sec:physicalrealism} aims to evaluate the extent to which the described approach achieves the goal of realism. Section \ref{sec:conclusion} presents concluding remarks and suggests future directions.

\footnotetext{\url{http://bf-astro.com/index.htm}}

\section{Related Work}
\label{sec:related}

%


Visualisation is an integral part of astronomical analysis, helping domain scientists to explore their data, identify problems or areas of interest for further analysis, as well as to convey important concepts and results during dissemination of their work \cite{Hassanetal11, SteffenEtAl14, Dykesetal18}. We first provide a brief contextual overview of visualisation approaches of simulated and mock data, and then focus on relevant research to the primary topic of the paper: multi-frequency and observed data reconstruction and visualisation.
 
Cosmological simulations allow astronomers to replicate physical processes to the best of their collective knowledge, simulating billions of years of evolution on timescales many orders of magnitude faster than real-time. Such simulations allow us to observe the behaviour of matter on the scale of the full evolution of the known universe in just a few days of computational processing using state of the art simulation codes and supercomputers \cite{Potteretal17, SpringelEtAl19}. The higher the resolution of a simulation, the easier it is to study discrete astronomical objects, such as single galaxies, as they evolve from simple clouds of matter to complex dynamic systems that contain hundreds of billions of stars \cite{Guedesetal11, NelsonEtal19, Pillepichetal19}. 3D visualisation in this context (e.g. \cite{Dolag09}) typically relies on the already existing 3D spatial structure from the simulation data. Our work, by contrast, is based on reconstruction and visualisation of 3D structure based on 2D observed images. We place emphasis on physical realism in our visualisation, by exploiting scientifically robust parameters where possible in our modelling and visualisation (discussed further in Section \ref{sec:physicalrealism}), other works have focused on achieving visual realism of scientific data through advanced rendering techniques, exploiting rendering software typically used in movie production to generate impressive cinematic visualisations of astrophysical data \cite{Naiman17,Borkiewicz19}. 

A related research area in Astronomy involves the creation of so-called \emph{mock galaxy catalogues}. Such catalogues, built from simulated data, consist of a structured collection of galaxies used to support development of scientific pipelines for the collection, validation, and analysis, of large scale observational surveys \cite{CrocceEtAl15}. This process includes \emph{mock imaging}, or the generation of synthetic observational images of simulated data. Such images can be used for the setup and tuning of observational instruments, and allow image-based comparison between results of theoretical simulation and physical observation for validation purposes (e.g. \cite{BottrellEtAl17, ZuHone18}). Tools such as SkyMaker \cite{Bertin09} and Phox \cite{Biffietal12} can be considered realistic visualisation tools that aim to generate 2 dimensional representations of simulated galaxies in specific frequency ranges, with many bespoke features such as replication of optical effects from telescopic instrumentation. In general our work targets 3D visualisation and exploration, as opposed to allowing quantitative comparison with observed images, and so we do not exploit this type of tool. 

There are a variety of works studying 3D visualisation of multi-dimensional astronomical data consisting of one or more spatial dimensions plus frequency. One approach is to consider the frequency domain as the third spatial dimension to use typical 3D visualisation methods, as in \cite{Lietal08}. Whilst effective for studying frequency domain data, this is not sufficient for a realistic representation in 3 spatial dimensions. In some cases, a velocity field can be derived from frequency data for a closer approximation to physical structure, as in \cite{Taylor15, Vohletal17, Punzo17}. Other approaches aim to reconstruct the proper 3D spatial structure for visualisation. Astronomical tomographic techniques for reconstructing 3D structure have been used to good effect, for example in \cite{Lee18}, who reconstruct and visualise areas of the Lyman-$\alpha$ forest, large clouds of gas that absorb Lyman-$\alpha$ radiation emitted from quasars. 

The work of \cite{Nadeau01} exploits a 3D modelling technique for the Orion nebula based on infrared and optical observational data \cite{Wen95} combined with volume rendering to visualise stars and emission nebulae. The work of \cite{Magnoretal04, Magnoretal05} presents a comprehensive modelling and visualisation pipeline for planetary nebulae, based on a novel \emph{Constrained Inverse Volume Rendering} technique to reconstruct gas distributions from a series of optical images; this technique further underpins the work of \cite{Hildebrandetal06}, who reconstruct and visualise spiral galaxy M\,81 based on optical and infrared images. The software tool Shape\cite{Steffenetal10} builds upon this modelling approach to provide an interactive construction and analysis tool for planetary nebulae, used in several related works focusing on reconstruction of planetary nebulae (e.g. \cite{Garcia12, Bandyopadhyay20}). The SlicerAstro \cite{Punzo17} extension to scientific visualisation software 3DSlicer \cite{Fedorov12} allows users to visualise (H\,{\sc i}) data in 3D, two spatial dimensions and a third velocity dimension. This is combined with interactive features for exploring such data and visualisation of tilted-ring based kinematic models (as discussed further in Section \ref{sec:observations}). 

The scope of our work is reconstruction and visualisation of observed galaxies based on multi-frequency data from observation catalogues. We present a methodology for galaxy reconstruction and visualisation that exploits a novel particle-based approach to reconstruct the 3D structure of spiral galaxies from multi-frequency observed images and kinematic models, utilising the high performance visualisation software Splotch for high quality visualisation. Previous works have covered a wide range of frequency bands, such as \cite{Lietal08}, but are focused on the visual exploration of data in those frequency bands rather than spatial reconstructions based on such data. The work of \cite{Hildebrandetal06} has similarly modelled spiral galaxies, however used fewer data sources and focused only on the stellar disk, achieving a less comprehensive galaxy view than our work. \cite{Steffenetal10} has developed a full interactive construction and analysis tool, however the focus is on other types of astronomical object such as astronomical nebulae, and the techniques do not directly apply to spiral galaxies. 

As will be described in following sections, our approach is able to: exploit a wider range of source data than related work, allowing us to incorporate an absorptive dust lane and the extended distribution of atomic hydrogen typically seen around spiral galaxies; utilise kinematic models to recover an approximation of 3D shape for spiral galaxies; and combine multiple objects into a single scene representing a group or cluster of galaxies. We also modify the Splotch software to better support such visualisations by extending the transfer function to treat emission and absorption fully independently to support dust occlusion in our visualisation.

\section{Observational structure and dynamics of galaxies}
\label{sec:observations}

This section presents a brief overview of the astronomical approach to inferring structure and kinematics of observed galaxies, which we will exploit for 3D reconstruction. We introduce two key domain-specific concepts, relating to the varied structure of galaxies observable across the electromagnetic spectrum (Section \ref{sec:structure}), and the concept of tilted-ring fitting for galactic hydrogen disks in spiral galaxies (Section \ref{sec:PMPV_3DVis_Kinematics}).

\subsection{Inferring the structure of observed galaxies}
\label{sec:structure}

In terms of physical structure, a typical spiral galaxy (like M\,83) consists of a bright stellar disk, an extended and warped gaseous hydrogen disk, a central elliptical bulge, and a dark-matter dominated halo. The stellar component is dominated by prominent spiral arms, which also contain large amounts of dust (see Fig.~1). A comprehensive understanding of this structure requires observations across multiple wavelengths. For example, the bright stellar body is typically captured at optical, infrared (IR) and ultraviolet (UV) wavelengths whilst cold atomic hydrogen gas (H\,{\sc i}), which is observed in spiral disk galaxies where it typically extends far beyond the bright stellar disk, is only observable with radio telescopes. Images of the stellar body observed at optical, IR, and UV wavelengths can inform us of different types of galactic population. Emission from the youngest, most massive, stars is mostly seen in UV, whilst older stars dominate optical observations, and different ranges of these spectra further segregate stellar populations by structure, age, and other physical properties \cite{Meureretal06, GildePazetal07}. IR emission in the 8$\mu$m range is typically used to identify and study interstellar dust \cite{Helouetal04}, which absorbs radiation other parts of the spectrum and re-emits it in IR.

To visualise galaxies of different sizes and types, an understanding of known relations can be beneficial. For example, the total \emph{dynamical} mass (a quantity derived from velocity observations) $M_{\rm dyn}$ of a disk galaxy scales with its rotational velocity and its radius $R$, i.e.  $M_{\rm dyn} \propto v_{\rm rot}^2 \times R$. Dwarf galaxies are low-mass, slow rotators, typically with a thick disk tending towards spheroidal shapes, while the most massive spiral galaxies have very large, thin disks. We distinguish between elliptical and disk galaxy components in Section \ref{sec:PMPV_3DVis_Model}, where different generating functions can replicate disk- or elliptical- shape galaxies.

For visualisation purposes, other galaxy properties such as the disk length and height of stellar and gas components can be determined by analysing a set of observations. It is often difficult to directly measure the height or disk thickness from 2D observed surface density maps, due to projection effects. However, in a study of nearby edge-on galaxies, \cite{Obrienetal10} found that the rotational velocity $v_{\rm rot}$ relates to the disk thickness $z_{\rm height}$; the faster galaxies rotate, the thinner their disks ($z_{\rm height} \propto 1/v_{\rm rot}$). As such, modelling the galaxy kinematics is essential in determining their 3D shape, as discussed in Section \ref{sec:PMPV_3DVis_Kinematics}.





\subsection{Kinematic modelling of warped disks}
\label{sec:PMPV_3DVis_Kinematics}



The hydrogen (H\,{\sc i}) disks of spiral galaxies are generally found to be (a) much more extended (by a factor 2--3) than the bright stellar disk and (b) warped as illustrated in Figure \ref{fig:3DVis_Generation_Kinematics_M83TiltedRing}. Such warps, which can range from a few degrees to tens of degrees, typically start at the edge of the bright stellar disk and extend outwards. To incorporate this warped shape in our reconstruction, we must approximate it in 3D.  

The de-facto approach for obtaining a best-fit shape for a H\,{\sc i} gaseous disk is via a \emph{tilted-ring analysis}, a long-standing means of investigating kinematic galaxy structure \cite{Rogstadetal74} based on modelling the H\,{\sc i} velocity field of galaxies for which we have high resolution observations (see, e.g., \cite{OhEtAl18}). A tilted-ring analysis generates a tilted-ring model,\ represented by the galaxy inclination ($i$), position angle ($PA$), thickness ($Z_0$), and rotational velocity ($v_{\rm rot}$), as a function of radius ($r$), and allows us to derive the 3D shape of H\,{\sc i}-rich spiral galaxies.



In this work, we utilise a tilted-ring model generated with the TiRiFiC software \cite{Jozsaetal07}. Our modelling approach requires one ascii text file per component consisting of the inclination angle, the position angle and the disk height as a function of radius; for future use we also include the rotational velocity as a function of radius. Such ascii files can be created from the output of any tilted-ring modelling software. For the purpose of 3D visualisation, tilted-ring models of regularly rotating galaxies are most useful. As the gaseous disks of galaxies are typically warped and much larger than the bright stellar body (factor 2--3), we require high resolution H\,{\sc i} 21-cm data for the TiRiFiC modelling.
This approach cannot be used for merging galaxies, where numerical simulations might allow to infer the 3D distribution of the stellar and gaseous galaxy components.

\begin{figure*} [ht]
\centering
\includegraphics[width=0.8\textwidth]{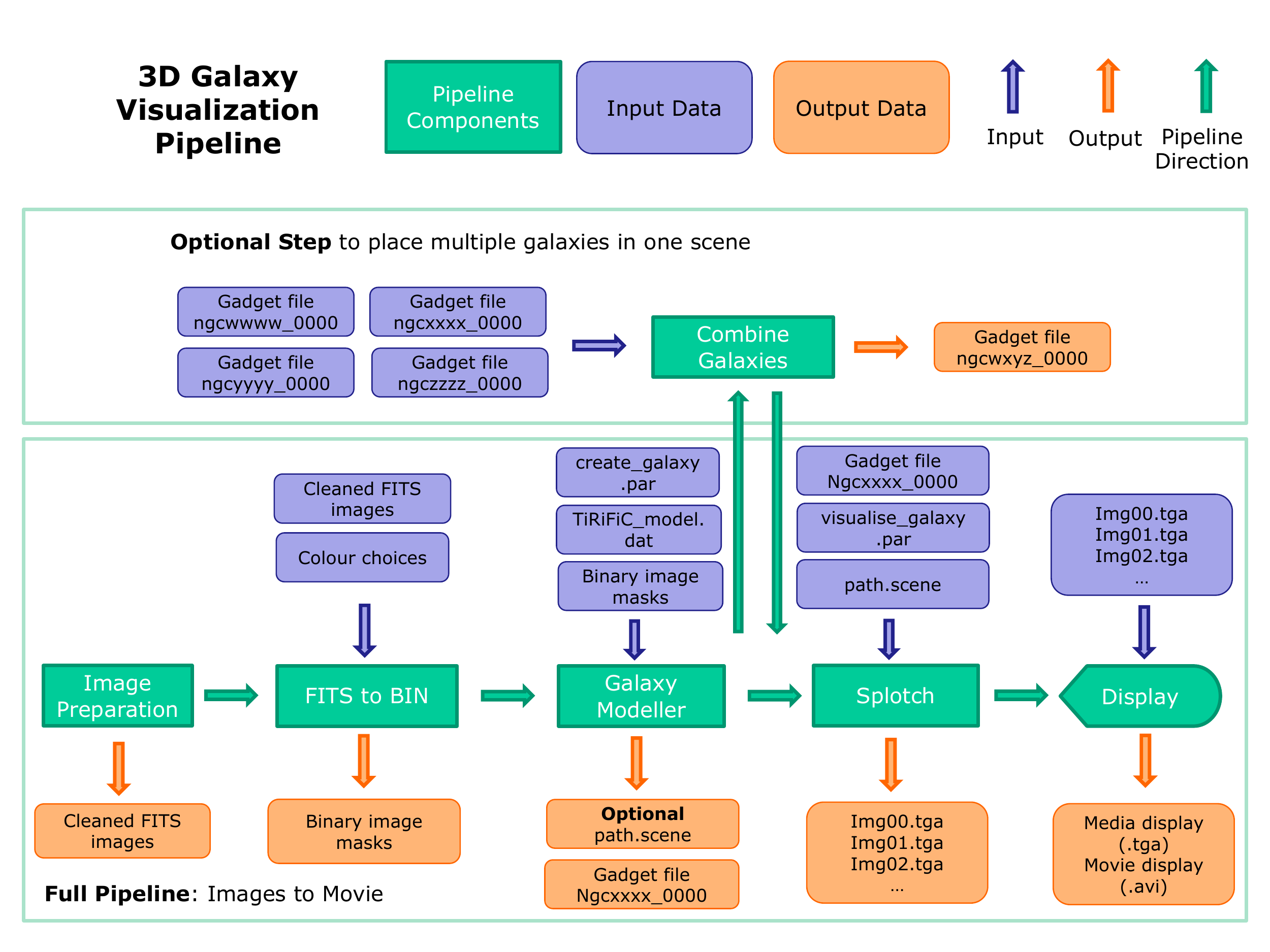}
\caption[The Galaxy Modelling Pipeline.]{A block diagram representation of the galaxy modelling pipeline starting from observed images of a specific galaxy, and resulting in visualised images of the reconstructed 3D galaxy model. Each of the stages, inputs, and outputs are described further in the text.}
\label{fig:GalaxyPipeline}
\end{figure*}

\section{Particle Based Modelling of Observed Galaxies}
\label{sec:modelling}

\subsection{Overview}

In this section we describe our approach for constructing a particle-based 3D galaxy representation based on observational data and a tilted-ring model as introduced in Section \ref{sec:observations}. Our representation consists of multiple galactic components:

\begin{itemize}
\item Stellar population
\item Diffuse hydrogen gas 
\item Dust 
\item Galactic bulge
\item Globular clusters surrounding the galaxy
\item Diffuse stellar, or dark matter, halo
\end{itemize}

We first collect and pre-process a set of observational source data and a tilted-ring model that will inform the density and spatial distribution of particles. Then, for each galaxy component, we construct a 3D distribution of tracer particles. In this context, a particle is considered a point source in 3D space with a series of inherent properties, such as a radius, a colour, and an intensity of light emission or absorption. The construction is implemented for each of our galactic components separately and is, where possible, based on the available observed and/or derived structural data. Components may also be tuned empirically using additional parameters which can be obtained from well known observed relations for the different galaxy types or components to present a more accurate representation of, for example, the thickness of the gas layer in the stellar disk (as in Section \ref{sec:PMPV_3DVis_Model}). Finally each of these tracer particles are coloured from observational images to reproduce the visual appearance as seen in the observations, and the galaxy components are combined to form the complete model.

The reconstruction and visualisation methodology is structured as a pipeline beginning with observed images of known galaxies, and ending with 3D visualisations of reconstructed, particle-based, galaxy models; the pipeline is illustrated by the block diagram shown in Figure \ref{fig:GalaxyPipeline}. This pipeline is used for both spiral and elliptical galaxies, with parameter files used to distinguish which inputs are required. The following sub-sections provide an overview of each stage of the modelling pipeline with associated inputs and outputs, illustrated via our case study galaxy M\,83. Section \ref{sec:visualisation} will then detail the visualisation process.

\subsection{Preparing the Source Data}
\label{sec:PMPV_3DVis_SourceData}

The first step of the pipeline is collection and preparation of the source data, which later will be used to inform the physical structure and colouring of different galaxy components; observed images are collected in the optical, UV, IR, and radio wavelengths, alongside 3D structural data from a tilted-ring analysis where possible. A variety of data is obtained to capture the diverse galactic populations as discussed in Section \ref{sec:observations}, exploiting existing survey data based on the availability for a specific galaxy.

In the case of M\,83, optical H$\alpha$- and $R$-band images are used from the Survey for Ionization in Neutral Gas Galaxies (SINGG) \cite{Meureretal06} and near- and far-$UV$ images from the Galaxy Evolution Explorer (GALEX) \cite{GildePazetal07} to inform the stellar component. To represent the structure of the gas component, radio interferometric H\,{\sc i} intensity maps at up to three different resolutions are created from the Australia Telescope Compact Array (ATCA) data. Most importantly, the H\,{\sc i} data is used to determine the extent, shape, and kinematics of galaxies, described in the next section. The ``Local Volume H\,{\sc i} Survey" \cite{Koribalski18} provides ATCA H\,{\sc i} data for nearly 100 nearby galaxies ($D < 10$ Megaparsec), which are publicly available\footnote{LVHIS page: www.atnf.csiro.au/research/LVHIS}. The dust distribution, which is to varying degrees already part of the observed stellar component, is informed by 8$\mu$m images (as in \cite{Helouetal04}) from the Spitzer Infrared Array Camera (IRAC) instrument \cite{Daleetal09}, which also provides improved structure to the spiral arms. Images are used from freely available from existing surveys, and so initial collection is quickly accomplished.

\begin{figure}
\centering
\includegraphics[width=0.8\columnwidth]{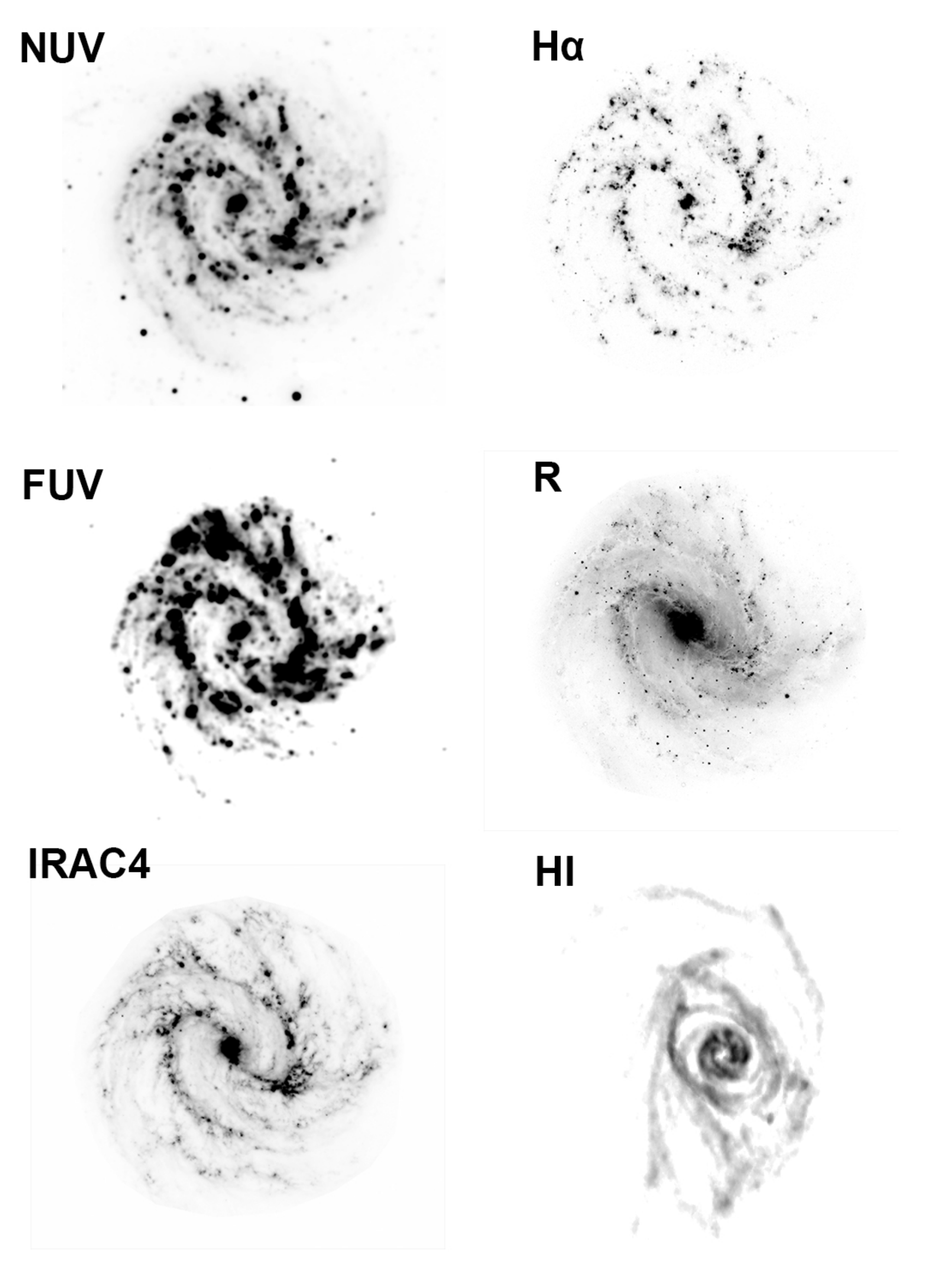} 
\caption[M\,83 UV, Optical, and IR Inner Disk Observations]{The cleaned FITS images of M\,83 used for reconstruction, in ultraviolet, optical, infrared, and radio bands. \emph{Top left:} GALEX near UV-band (UV stellar distribution). \emph{Middle left:} GALEX far UV-band (UV stellar distribution). \emph{Bottom left:} SPITZER IRAC 8$\mu$m-band (Dust distribution). \emph{Top right:} SINGG H$\alpha$-band (Optical stellar distribution). \emph{Middle right:} SINGG $R$-band (Optical stellar distribution). \emph{Bottom right:} ATCA radio-band at medium resolution (Combined with other resolutions for diffuse (H\,{\sc i}) distribution).}
\label{fig:M83_InnerDisk_Observations}
\end{figure}
\begin{figure}
\includegraphics[width=\columnwidth]{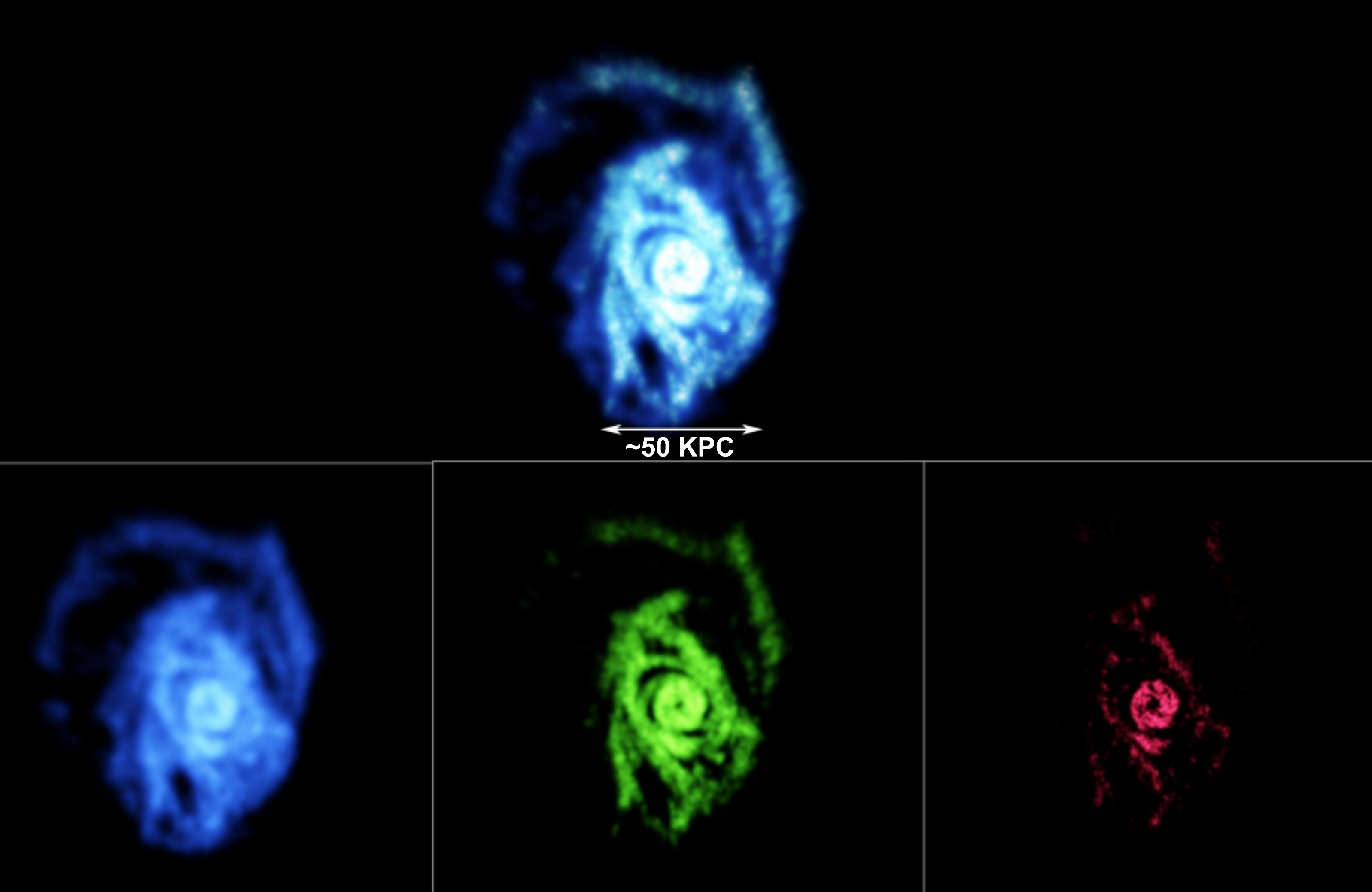} 
\caption[M\,83 Multi-resolution Radio Outer Disk Observations]{M\,83: cleaned, preprocessed, and coloured FITS images of the warped outer disk in high, medium, and low spatial resolutions. The differing resolutions are defined by a weighting applied when extracting images from the raw data, allowing to tune the extent of detected emission. \emph{Bottom left:} Low, capturing the diffused emission. \emph{Bottom Middle:} Medium, capturing most of the details. \emph{Bottom Right:} High, capturing the sharp features. \emph{Top middle:} Combined resolution, approximating the high dynamic range of the original data.}
\label{fig:M83_OuterDisk_Observations}
\end{figure}

The multi-wavelength images are then pre-processed and organised as a set of FITS \cite{Penceetal10} files. The FITS (Flexible Image Transport System) format is an open standard commonly used in astronomy for storage, transmission and processing of data, typically in the form of 2D or 3D images, or tables. Pre-processing is a more involved task than initial collection, requiring an experienced astronomer to clean, normalise, orient, and scale images. For example, optical images of galaxies contain foreground stars and background galaxies that need to be removed (or \emph{cleaned}) using domain-specific software (e.g. \cite{Bertin96}). This stage is important as extraneous sources will introduce incongruous artefacts in the final 3D model. The normalisation, orientation, and scaling of images is required to ensure each galaxy component is modelled accurately relative to each other component - this task is carried out with the same rigour as for typical image analyses. Image intensity ranges are then clipped to reduce the dynamic range of pixels. Source images are typically FITS formatted with pixels of high dynamic range, these cannot be accurately represented in the colour depth (or radiometric resolution) of images typically used in 3D visualisation. An experienced astronomer must choose suitable maximum and minimum cutoff values to represent the respective galaxy component, a maximum that is too high will result in over-saturation, an overpopulated model, whilst a minimum that is too low will result in under-saturation, a sparsely populated model. Figures \ref{fig:M83_InnerDisk_Observations} and  \ref{fig:M83_OuterDisk_Observations} show the collection of images used for M\,83 after pre-processing.

\begin{table}
\centering
\begin{tabular}{rrrrr} 
\hline
\multicolumn{1}{l}{$r$} & \multicolumn{1}{l}{$v_{\rm rot}$} & \multicolumn{1}{l}{$Z_0$} & \multicolumn{1}{l}{$i$} & \multicolumn{1}{l}{$PA$}  \\ 
\hline
0                    & 150                 & 29.9               & 23.7                 & 227                \\
172                  & 196                 & 29.9               & 23.7                 & 227                \\
343                  & 200                 & 29.9               & 24.7                 & 226                \\
515                  & 195                 & 29.9               & 34.2                 & 219                \\
686                  & 192                 & 29.9               & 44.8                 & 209                \\
858                  & 190                 & 29.9               & 50.6                 & 201                \\
1030                 & 189                 & 29.9               & 51.0                 & 196                \\
1200                 & 189                 & 29.9               & 50.8                 & 192                \\
1370                 & 189                 & 29.9               & 50.1                 & 188                \\
1540                 & 189                 & 29.9               & 48.9                 & 188                \\
1720                 & 190                 & 29.9               & 43.3                 & 188                \\
1890                 & 189                 & 29.9               & 36.9                 & 182                \\
2060                 & 189                 & 29.9               & 30.6                 & 174                \\
2230                 & 189                 & 29.9               & 24.9                 & 165                \\
5230                 & 189                 & 29.9               & 24.9                 & 165                \\  
\hline
\end{tabular}
\caption[The tilted-ring model values for M\,83.]{A minimal list of tilted-ring model values for M\,83, visually represented in Figure \ref{fig:3DVis_Generation_Kinematics_M83TiltedRing}. Each concentric ring is defined by a radius $r$, rotational velocity $v_{\rm rot}$, thickness $Z_0$, inclination angle $i$, and position angle $PA$. For brevity only 15 representative rings are shown, the full model contains $\sim$300 rings and was generated by Peter Kamphuis using the TiRiFiC tilted-ring fitting software \cite{Jozsaetal07}.}
\label{tab:3DVis_Generation_Kinematics_TiltedRingValues}
\end{table}

The cleaned FITS images are then input to the \emph{FITS to BIN} pipeline stage, consisting of a preprocessing script that extracts, clips, and colours the FITS image pixels. The script is given, as input, colour values as RGB triplets per input image, these colour values are combined with a per-pixel intensity and scaled to 8 bit RGB pixels for output as a binary image mask for the galaxy modeller. Clipping values may also be provided here to further clip minimum and maximum intensities to tune the saturation of the model. Colour values are chosen typically to highlight the different components, and if possible to match existing composite observed images.

\begin{figure}
\centering
\includegraphics[width=\columnwidth]{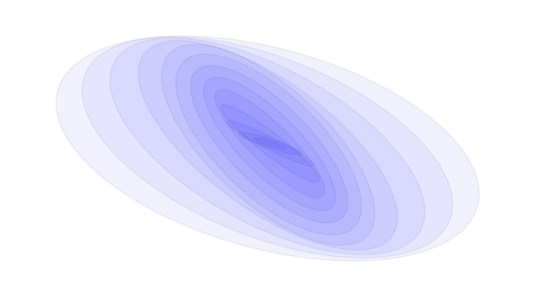}
\caption[A visual representation of the M\,83 tilted-ring model.]{A visual illustration of the TiRiFiC tilted-ring model for M\,83, based on the data in Table \ref{tab:3DVis_Generation_Kinematics_TiltedRingValues}.}
\label{fig:3DVis_Generation_Kinematics_M83TiltedRing}
\end{figure}

A tilted-ring model is used to inform the 3D structure of the galaxy disk. For M\,83, we use a model generated with the TiRiFiC software, which provides a set of concentric ellipses of varying radii $r$ and thickness $Z_0$, with $i$, $PA$, and $v_{\rm rot}$ values as shown in Table \ref{tab:3DVis_Generation_Kinematics_TiltedRingValues}. This model, visualised in Figure \ref{fig:3DVis_Generation_Kinematics_M83TiltedRing}, effectively describes the warped disk structure of M\,83, and is also provided as input to the galaxy modeller.

\subsection{Building the 3D Model}
\label{sec:PMPV_3DVis_Model}


This section describes the construction of the spatial distribution of particles for each galactic component via the galaxy modeller pipeline stage. The modelling is implemented in C++, using key-value ASCII text files for parameter inputs. Throughout this section, coordinates $(x,y,z)$ are used relative to the galactic plane, such that $x,y$ are in-plane and $z$ is axial (normal to the galactic plane). 

The stellar, diffuse gas, and dust particle populations of the disk are based on the tilted-ring kinematic model described in Section \ref{sec:PMPV_3DVis_Kinematics}. Firstly, a 3 dimensional distribution of $N_p$ \emph{seed} particles is generated in an unweighted random way on the galactic plane at $z=0$, defined by the concentric tilted-ring model. This distribution of particles in object space is projected onto the image plane of the corresponding observational image, $Obs_\lambda$. Each seed particle $P$ is assigned an intensity $I_s$ as a function of the intensity $I_p$ of the image pixel with which it intersects, along with a scalar or three-component $RGB_s$ colour value defined by the binary image masks generated in Section \ref{sec:PMPV_3DVis_SourceData}. An intensity threshold is defined as $I_t$, defaulting to $0$, and all seed particles with $I_s < I_t$, are culled, i.e. those that lay in an area of the disk with no emission seen in image $Obs_\lambda$. This creates a flat disk in the warped galactic plane with a distribution of seed particles matching the observed image. 

The disk thickness is constructed by generating point cloud distributions around the seed particles. For each of the remaining particles, a point cloud of size $N_{star}$ is generated according to a Gaussian distribution defined by $\sigma_{star}$, and $N_{star}$. Each spawned particle inherits the intensity $I_s$ and colour values $RGB_s$ of the seed particle, and is further characterised by a \emph{smoothing length} $h$, a term from astrophysical simulation equivalent to the average inter-particle distance and used during visualisation in Section \ref{sec:visualisation}. The number of particles, and their displacement from the galactic plane, are defined by $N_{star}$ and $\sigma_{star}$ respectively and determine the thickness of the disk. These can be calculated using one of three available models:

\begin{enumerate}
\item A Gaussian thickness defined by tunable input parameter $\sigma_{xyz}$, where particles are distributed with $\sigma_{star} = \sigma_{xyz}$ and $N_{star}$ is scaled with $I_s$. As such, brighter pixels of the image are represented by more points across a larger spatial volume. The scaling of $N_{star}$ with $I_s$ is defined as the inverse of the function defining $I_s$ from $I_p$, such that $N_{star} \cdot I_s \approx I_p$. 
\item Radial ($H_r$) and axial ($H_z$) parameters are provided, along with tunable input parameter $\sigma_{xy}$. $N_{star}$ is scaled proportionally to $exp(-R/H_z)$ (where $R$ is radial distance of the particle to the center of the disk), and the particles are distributed with $\sigma_{xy}$ in the radial plane and $\sigma_z$ axially, representing a tapered disk structure.
\item Thickness is scaled to fit to known measurements of flared disk thickness in edge-on galaxies, following the measurement of $FWHM_{z,g}$ (gas layer thickness) as presented in Figure 25 of \cite{Obrienetal10}.
\end{enumerate}

In general, the appropriate model is chosen to match observed properties. Model 1 may be used as a heuristic approach approximating a Gaussian thickness. Model 2 provides a disk thickness matching generally observed tapered galactic disks, whilst Model 3 provides a thickness matching observations of flared H\,{\sc i} disk galaxies. In the case of M83, Model 2 is used for the stellar disk components, whilst Model 3 is used for the flared gaseous H\,{\sc i} disk. The model may be chosen per-component in the parameter file passed to the galaxy modeller stage of the pipeline shown in Figure \ref{fig:GalaxyPipeline}. 

As can be seen in high quality edge-views of spiral galaxies, for example that shown in Figure \ref{fig:M83_NGC4565} (right), dust lanes can have complex morphologies with cloud and filament-like structures, and can be very well-defined against the bright stellar background. To reflect this, the thickness of the dust component is implemented using a slightly different scheme. The point clouds generated around seed particles are distributed using one of two models:

\begin{enumerate}
\item Dust filaments are approximated via assigning velocities to particles randomly distributed around the linear velocity defined by the TiRiFiC model. A random walk is traced, generating a new particle at each step, with a gravitational factor applied such that filaments are constrained near to the galactic plane.
\item As with the previous item, however particle velocities are distributed around the rotational velocity of the disk, constraining filaments to the same rotation as the disk.
\end{enumerate}

As discussed further in Section \ref{sec:physicalrealism}, the dust thickness model is addressed with a more heuristic approach than the disk, and so the appropriate model is chosen for visual effect. 

The galactic bulge is more simply defined as a Gaussian distribution of points defined by input parameters $\sigma_x$, $\sigma_y$, $\sigma_z$, and $N_{bulge}$, and coloured white ($RGB_{bulge} = [1,1,1]$). The bulge can optionally be rotated to conform to the tilted-ring kinematic model. Finally, the globular cluster, and diffuse stellar/dark matter halo components are available in the pipeline for visual effect in a multi-galaxy scene. They are generated using the same algorithm implemented for the galactic bulge, with the caveat that the position of the centre of each globular cluster is user defined. Neither of them is used in this work. 

For each component, the particle counts ($N_p$, $N_{star}$, $N_{bulge}$) are provided as user inputs, whilst the final number of particles depends on the input image (a more saturated image results in production of more particles, as there is more emission to emulate). Currently these initial values are tuned empirically based on visualisation of the resulting model using an Earth-based view and comparison to observations; too many particles result in an over-saturated model where individual features are difficult to distinguish, whilst too few result in an under-saturated model where features are missing or not well resolved. 

All the components described above can be combined to create complex galaxies, like M\,83. However, each different component can also be used to model simpler objects, like elliptical or dwarf galaxies. The former can be represented as a stellar ellipsoid, or nested ellipsoidal distributions representing a bulge and extended star distribution, characterised by different values for axis and star density. The latter may represent various types of small galaxies that can have either a disk-like structure, in which case the \emph{stellar population} model is used, or a spherical shape, in which case the \emph{galactic bulge} model is used. For both types of small galaxy, a hydrogen gas encompassing cloud can be present, which may be added using the \emph{diffuse gas} component. We have used these models in Section \ref{sec:results} to construct and visualise the M\,83 local group of galaxies. 


Finally, cleaned and pre-processed observational images (as described in Section \ref{sec:PMPV_3DVis_SourceData}) together with the binary image masks are used to provide $RGB_s$ colour values to paint the galaxy population. In case where images are not available (typically the case with bulges or globular clusters) a colour can be chosen manually to visually distinguish each of the galactic components, and if possible to resemble existing recognisable false-colour composite images.
Each of the components is generated sequentially, and the full galaxy representation is then composited and written as one of a series of commonly available file types in astronomy (CSV, Gadget, or HDF5). All of the parameters can be provided though an input key-value text file shown as \emph{create\_galaxy.par} in Figure \ref{fig:GalaxyPipeline}.

\subsection{Combining Galaxies}
\label{sec:localgroupreconstruction}

Once several galaxies have been modelled, they can be combined within the same scene to render, for example, a group of neighbouring objects bound by gravity in the same system. This could represent a local group of galaxies made of several components, like the Milky Way and the Magellanic Clouds, or a cluster of galaxies, like the Coma or the Virgo clusters, composed of hundreds to thousands of galaxies. 

The various components are placed in the position identified by their known astronomical coordinates, converted to a Cartesian reference frame for compatibility with typical visualisation tools. Their size may be arbitrarily scaled for visibility; the distance between galaxies is typically orders of magnitude bigger than their size, such scaling is then required for concurrent visualisation of multiple objects with a moving camera. The scaling may be the same for all elements in the scene, to preserve the size ratio between objects, or specific to each member to highlight given elements of the group. In both cases, however, the relative positions and orientations of galaxies are preserved, such that the resulting image forms a realistic, although locally magnified, representation of the actual system. 
Galaxy combinations are merged after construction, creating a single dataset that may be visualised. An application of such combination is presented in Section \ref{sec:results}, describing the M\,83 galaxy local group.

\section{3D Galaxy Visualisation}
\label{sec:visualisation}

This section describes the visualisation process for a constructed galaxy model. We utilise the astronomical visualisation tool Splotch, which is well-suited to our purpose as it is designed for particle-based astronomy data and supports multiple \emph{species} of particle, natively supporting the multi-component particle-based structure of our galaxy models. The software, with which the authors have significant development experience, is open-source which allowed us to modify the underlying rendering algorithm to better support our galaxy models as described in Section \ref{sec:splotch}. Future work (as discussed in Section \ref{sec:conclusion}) will also benefit from the high-performance parallel nature of Splotch as we scale up from single galaxies to large groups.

\subsection{Splotch}
\label{sec:splotch}

Splotch is a high performance and scalable scientific visualisation package designed for particle-based astronomical datasets, with implementations for a variety of hardware platforms  \cite{Dolagetal08,Jinatal10,RiviEtAl14,Dykesetal17}. Splotch has been utilised in the past for a variety of types of visualisation, including illustrating numerical simulation results in academic talks, scientific communication and outreach through generation of movies for planetariums, and supplementing scientific analysis through visualisation in the context of theoretical virtual observatories \cite{Ragagnin17, Dykesetal18}.

The Splotch software is written in C++, with minimal dependencies beyond those for parallel models (e.g. OpenMP\footnote{\url{https://www.openmp.org/}}, CUDA\footnote{\url{https://developer.nvidia.com/cuda-zone}}, the Message Passing Interface (MPI)\footnote{\url{https://www.mpi-forum.org/}}) and specific file I/O. Splotch is motivated by cosmological N-body and SPH point-like simulation data, which describe fluid flow using tracer particles that are spread across a 3D domain using a kernel such as the $B_2$-Spline \cite{Monaghan85}, defined such that particles overlap with a set number of neighbours. 

\begin{figure} [ht]
\begin{center}
\includegraphics[width=\columnwidth]{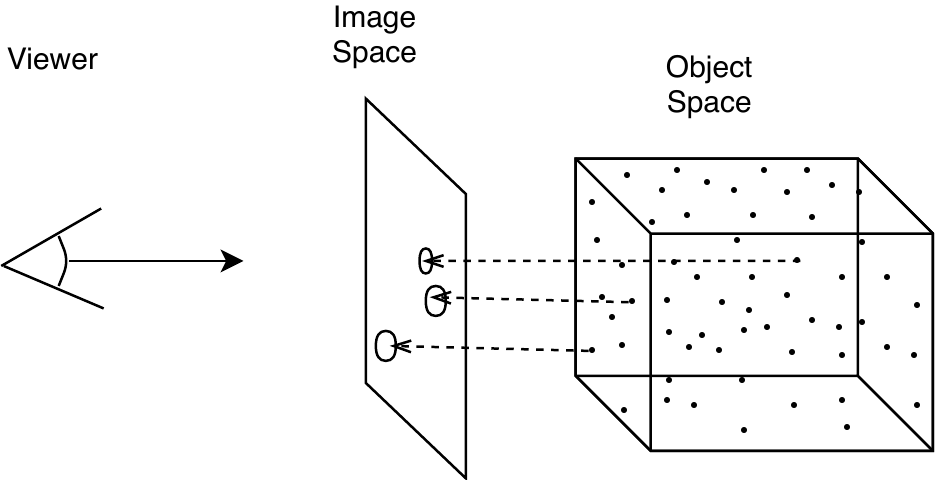} 
\caption[Volume splatting]{Volume splatting for particle data: particles are projected from object space to image space, and \emph{splatted} across the image using a footprint function, typically a Gaussian kernel.}
\label{fig:VolumeSplatting}
\end{center}
\end{figure}

The rendering method of Splotch is an implementation of volume splatting \cite{Westover92}. First each data element is transformed relative to a viewpoint, a parallel or perspective projection applied, and then the contribution of each element to line-of-sight rays cast from image pixels is computed using a ``splatting" kernel, summarised in Figure \ref{fig:VolumeSplatting}. In Splotch, each data element is represented by a particle, and a simplified emission and absorption optical model (\cite{Max95}) is used to define each particle's contribution to the rays as follows, starting from the radiative transfer equation in differential form:

\begin{equation}
\label{eq:RadTransfer_FormalDifferential}
   \frac{dI(x)}{dx}=(E_p-A_p I(x))\rho_{p}(x)
\end{equation}

\noindent
$x$ is the coordinate along the line of sight, $I$ is the intensity at position $x$, $E_p$ and $A_p$ are the emission and absorption coefficients of particle $p$. In this form, both $E_p$ and $A_p$ directly rely on $\rho_p(x)$, which defines the contribution to matter density interpolated from particle $p$, and is defined using a Gaussian distribution:

\begin{equation}
\label{eq:RadTransfer_SplotchSmoothing}
   \rho_p(x)=\rho_{0,p}\exp(-\mid\mid x - x_p \mid\mid ^2/\sigma_p^2)
\end{equation}

\noindent
with $x_p$ representing the particle coordinates, and $\rho_{0,p}$ and $\sigma_p$ being the mass density and the radius of the particle respectively. For a more convenient compact support, the distribution is truncated at $\chi \cdotp \sigma_p$, where $\chi$ is a suitably defined factor typically chosen such that $\chi \cdotp \sigma_p \approx h$; where $h$ is the intrinsic \emph{smoothing length} of the particles (e.g. as described in \cite{Koda99}). Due to this relation, the particle radius is commonly referred to as the smoothing length in this document and the referenced Splotch publications. As noted in \cite{Dolagetal08}, the $B_2$-Spline typical for SPH particles is very similar in shape to the Gaussian distribution used here. 

Following on from Equations \ref{eq:RadTransfer_FormalDifferential} and \ref{eq:RadTransfer_SplotchSmoothing}, the contribution of a single particle to a ray is defined as:

\begin{equation}
\label{eqn:RadTransfer_SplotchIntensity}
I_{after} =  (I_{before} - E_p/A_p)\exp(-A_p\int\limits_{-\infty}^{\infty} \rho_p(x)\ dx) + E_p/A_p
\end{equation}

For simplicity, the frequency dependency of the intensity is not included in Equations \ref{eq:RadTransfer_FormalDifferential} to \ref{eqn:RadTransfer_SplotchIntensity}; however, such a dependency does exist. The transfer function of Splotch supports emission, and absorption as a function of emission, in three frequencies corresponding to colors  R, G and B, and referred to hereafter as floating point triplets $E_{RGB}$ and $A_{RGB}$ respectively. Furthermore, each particle has an intrinsic \emph{type} property, used to distinguish particle species (for example gas, stars, and black holes). The transfer function can additionally be customised per type, to support an arbitrarily large range of functions each with three frequency outputs.

This optical model requires the particle data to be sorted back-to-front with respect to the viewer, such that absorption and emission can be integrated in the correct order. Splotch also supports a further simplification, an assumption that $E_p = A_p$ removes the need for order dependent rendering, and acts as a high performance approximation for highly diffuse or optically thin material (e.g. intergalactic medium), or extremely compact and bright material (e.g. stars), both of which are common in astrophysical simulation. As such, a flag can be set to indicate $E_p = A_p$, neglecting sorting before rendering, however this mode is not used for the model visualisation in Section \ref{sec:model_vis} due to the described treatment of stellar dust. 

We extended the Splotch rendering software to treat emission and absorption coefficients independently, rather than treating absorption as a function of emission, to support an absorptive galactic dust component. The extended Splotch code allows the user to provide per particle coefficients for both emission and absorption, in three frequencies, from their source data. These are included in the parallelised radiative transfer computation in the rendering kernel at a cost of 3 additional floating point fields per particle, or a 30\%  increase in memory consumption during execution. To support this independent absorption coefficient, an algorithmic extension is also needed to account for the lower limit of absorption, i.e. for a purely emitting particle or one with negligible absorption, $ A_p \approx 0 $. In this case, Equation \ref{eqn:RadTransfer_SplotchIntensity} is replaced with a more simple emission-only optical model:

\begin{equation}
\label{eqn:Radtransfer_AbsorptionLimit}
   I_{after} = I_{before} + E_p\int\limits_{-\infty}^{\infty} \rho_p(x)\ dx
\end{equation}

The additional fields and extended rendering algorithm are implemented with C pre-processor commands allowing them to be switched off at compile time. This supports a quick reversion to the simplified rendering model, in the case where there is no sensible means of modelling the absorption.

\subsection{Model Visualisation}
\label{sec:model_vis}

\begin{figure*}
\centering
\begin{tabular}{cc}
\medskip
(a)\raisebox{-.9\height}{\includegraphics[width=0.8\columnwidth]{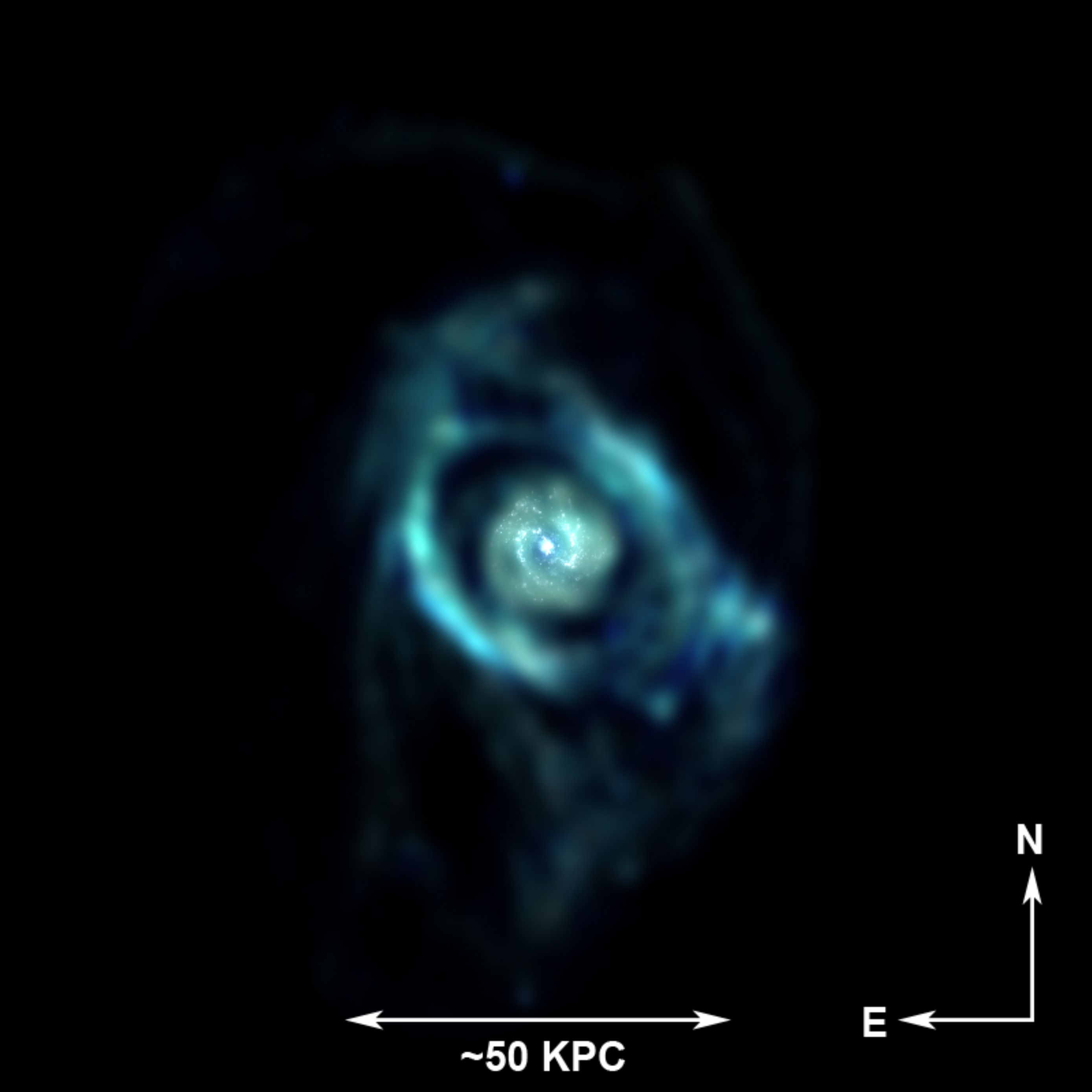}} & 
(d)\raisebox{-.9\height}{\includegraphics[width=0.8\columnwidth]{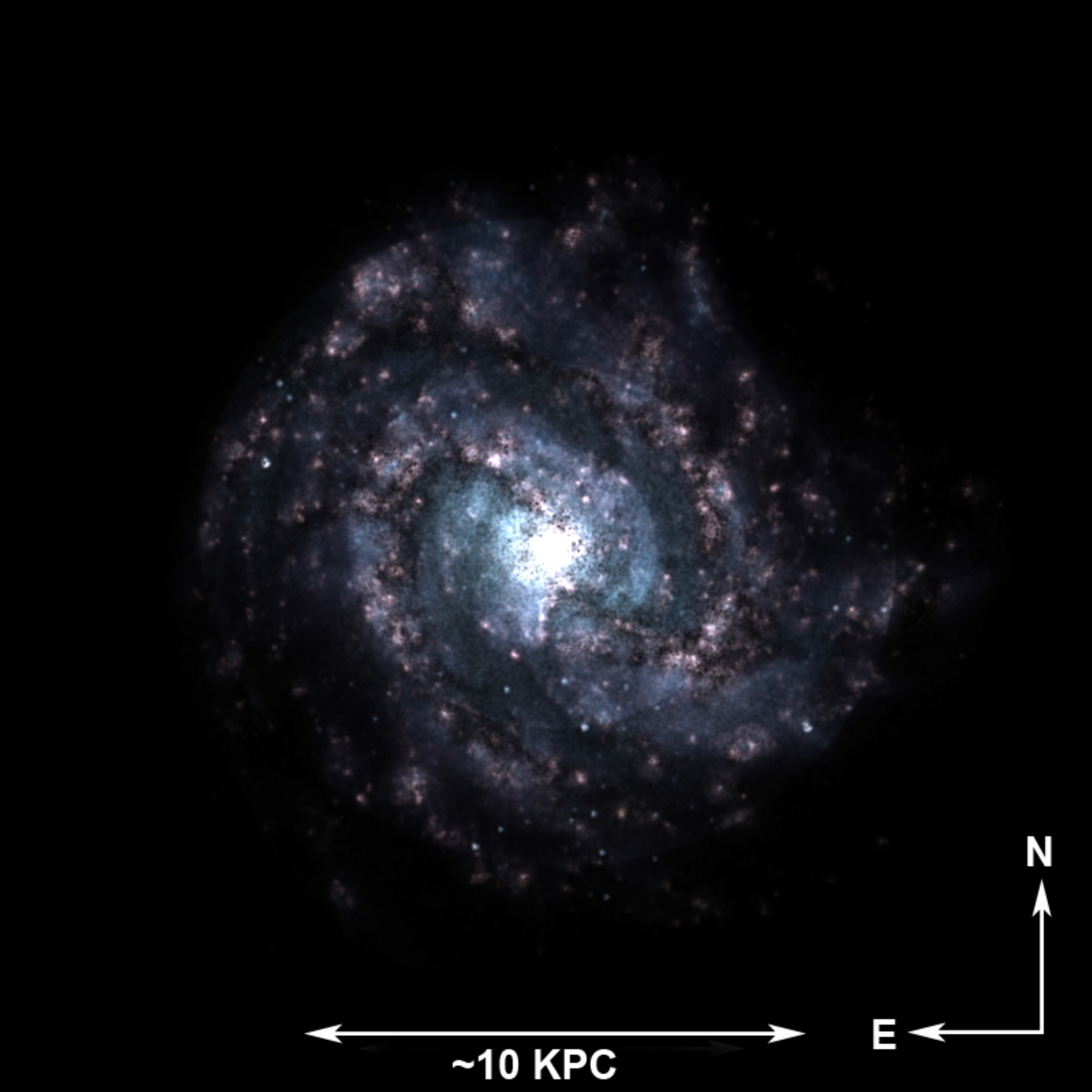}} \\
\medskip
(b)\raisebox{-.9\height}{\includegraphics[width=0.8\columnwidth]{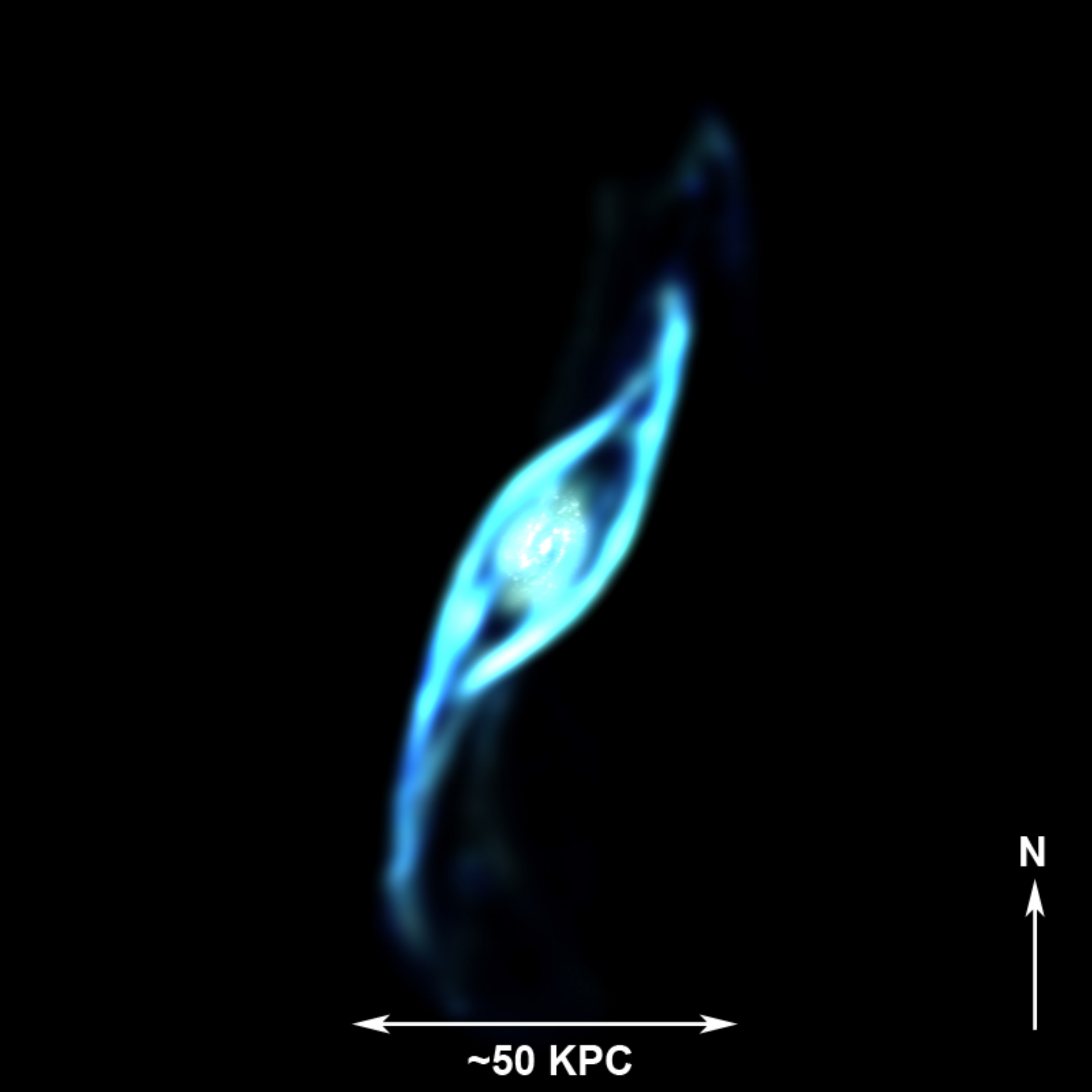}} & 
(e)\raisebox{-.9\height}{\includegraphics[width=0.8\columnwidth]{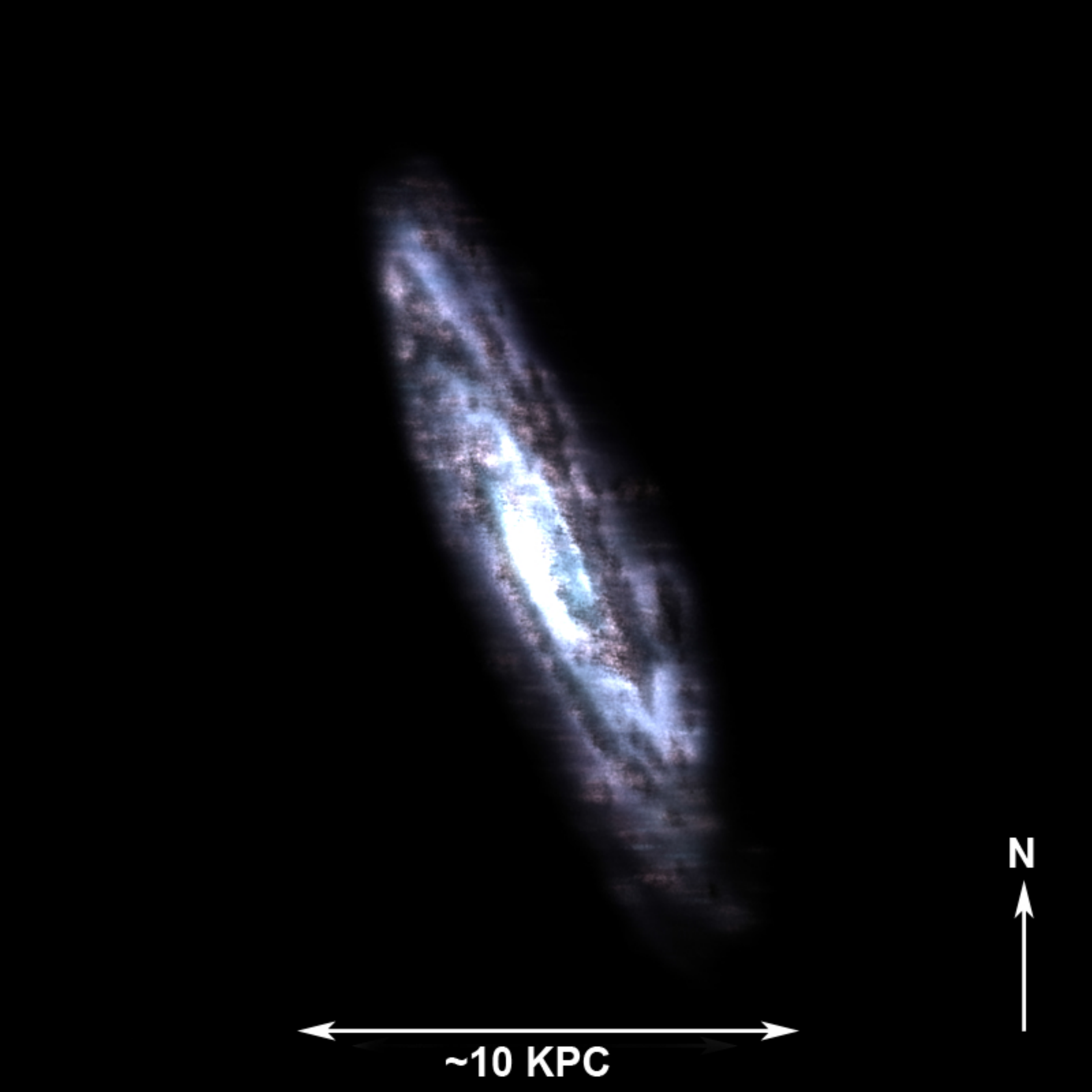}} \\
\medskip
(c)\raisebox{-.9\height}{\includegraphics[width=0.8\columnwidth]{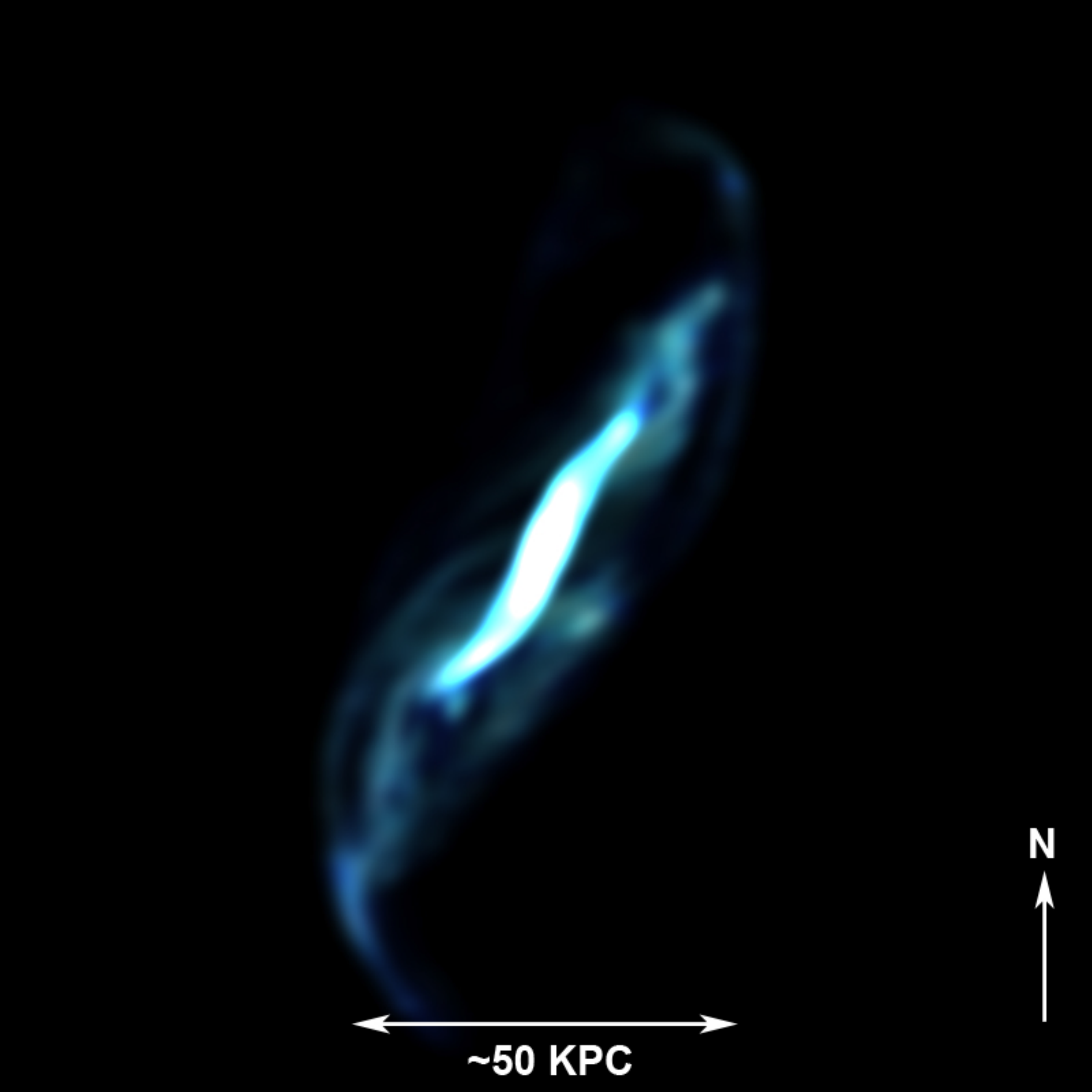}} & 
(f)\raisebox{-.9\height}{\includegraphics[width=0.8\columnwidth]{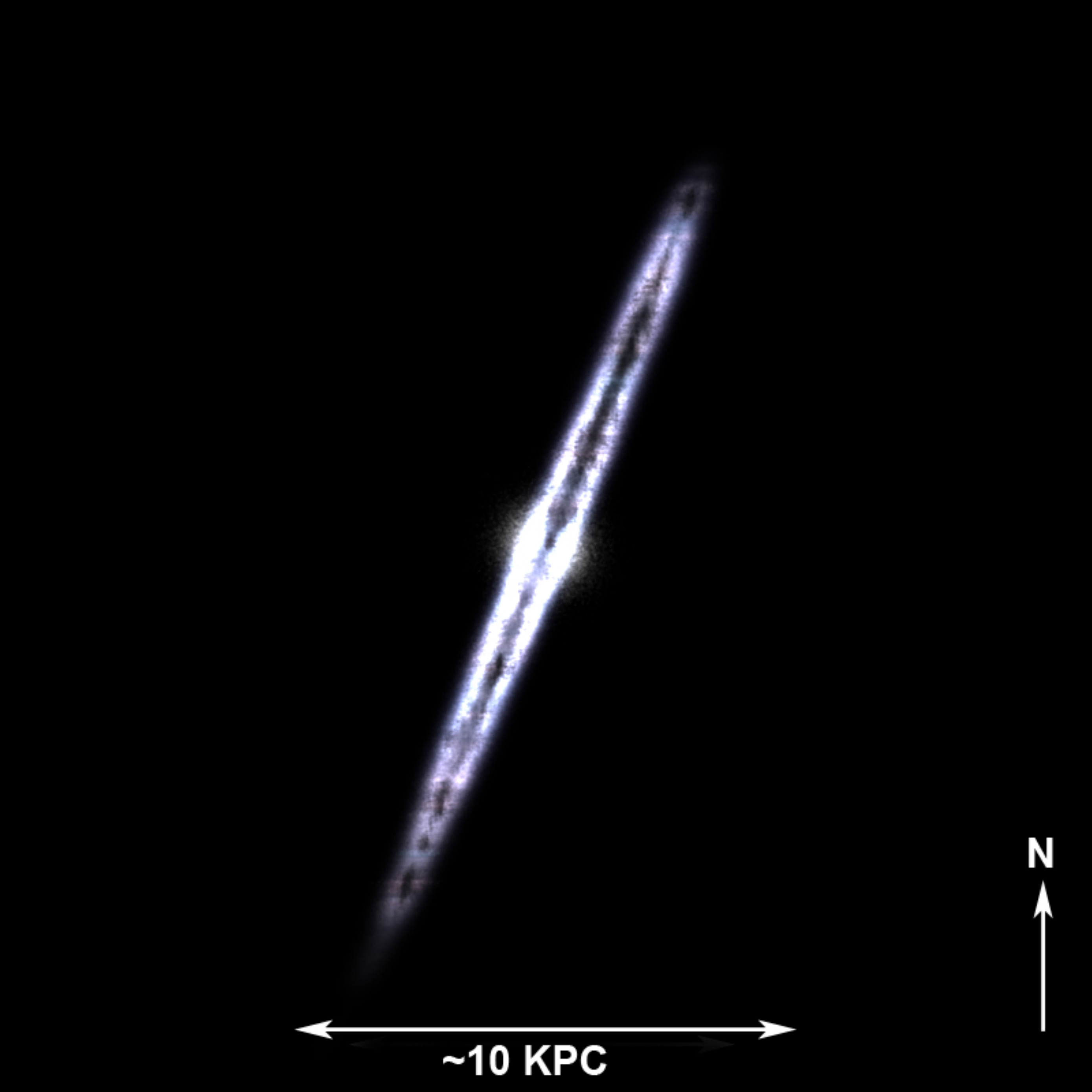}} \\
\end{tabular}{}

\caption[M\,83 visualisation results, face-on, angled, and edge-on.]{Example visualisation outputs for the M\,83 galaxy, split into \emph{far} and \emph{close} (\emph{left} and \emph{right}, respectively), showing face-on, angled, and edge-on views (\emph{top} to \emph{bottom}, respectively). The far images \emph{(a, b, c)} include the extended gaseous disk ($\sim$100 kpc diameter), illustrating the large warped structure. The close images \emph{(d, e, f)} zoom into the stellar disk and have the H\,{\sc i} removed to more closely resemble an optical composition ($\sim$17 kpc diameter), and show clearly the absorptive dust lanes, tapered disk and stellar bulge. The added image scales give a smaller impression, as the edges of the galaxy fade into darkness, implying a slightly smaller structure.}
\label{fig:M83_Visualisation_results}
\end{figure*}

Splotch takes as input: the data file written by the Galaxy Modeller (e.g. \emph{NgcXXXX\_0000} as shown in Figure \ref{fig:GalaxyPipeline}), a key-value parameter file (\emph{visualise\_galaxy.par}) describing the scene configuration, and optionally a scene file (\emph{path.scene}) which can be used to describe a set of scene configurations for a movie.

For each of the galactic components, which are treated as separate particle species in Splotch, a series of tunable visual parameters are available. The smoothing length $h$ of particles can be scaled using a \emph{size} parameter, and the intensity $I$ can also be scaled using a \emph{brightness} parameter. 

The emission and absorption coefficients of each component are defined by the intensity $I$ of the particle, retained as $I_s$ from Section \ref{sec:PMPV_3DVis_Model}, such that the emissivity of the particles is directly related to the observed intensities of the galactic component to which the particle belongs. The emission coefficient in each frequency of the final image $E_{RGB}$ is defined as $I_s \cdot RGB_s$. The absorption coefficient $A_{RGB}$ is defined as $I_s \cdot RGB_A$, where $RGB_A$ is a three component absorption profile provided as a lookup table during transfer function evaluation.

In the example case for M\,83, there are 5 galactic components included in the visualisation, which are informed by observations as illustrated by Table \ref{tab:M83_Components}. The stellar distribution and diffuse hydrogen gas are treated as coloured emissive sources, where $E_{RGB} = I_s \cdot RGB_s$ and $A_{RGB} = 0$. The bulge component is treated as a white fully emissive source, where $E_{RGB} = I_s \cdot [1,1,1]$ and $A_{RGB} = 0$. The dust is defined as a grey fully absorptive source, where $RGB_s = [0,0,0]$ and $A_{RGB} = I_s \cdot [1,1,1]$. 

\begin{table}[tbp]
\small\sf\centering
\begin{tabular}{llllll}
\hline 
\noalign{\smallskip}
\textbf{Galactic Component} & \textbf{Source Imaging Data} \\ 
\hline
Stellar distribution (Optical) & SINGG H$\alpha$- and $R$-band \\
Stellar distribution (UV) & GALEX Near and Far UV \\
Diffuse hydrogen gas & ATCA H\,{\sc i} at three resolutions  \\
Dust & Spitzer IRAC 8$\mu$m  \\
Galactic Bulge & None  \\ 
\hline
\end{tabular}
\caption[Galactic components mapped to source images and visualised quantities.]{A mapping demonstrating the relationship of galactic components used for M\,83 to source image data.}
\label{tab:M83_Components}
\end{table}

\section{Reconstruction and Visualisation of M\,83 and its local group}
\label{sec:results}

Figure \ref{fig:M83_Visualisation_results} demonstrates the results of applying our reconstruction and visualisation methodology to the M\,83 galaxy. The images are split into \emph{far} and \emph{close} (\emph{left} and \emph{right} respectively), showing face-on, angled, and edge-on views (\emph{top} to \emph{bottom} respectively). The far images include the H\,{\sc i} extended gaseous disk, illustrating the large warped structure which is not easily discernible for an astronomer viewing the observed H\,{\sc i} images (Figure \ref{fig:M83_OuterDisk_Observations}). The maximum extent of the H\,{\sc i} region is $\sim$100 kilo-parsecs (kpc) \cite{Koribalski18}. In contrast the close images show the inner stellar disk, which is $\sim$13 kpc diameter \cite{Thilkeretal05}, and have the H\,{\sc i} removed to more closely resemble an optical image (such as Figure \ref{fig:M83_NGC4565} \emph{left}). These close images highlight the spiral arm structure, absorptive dust lanes, tapered disk and stellar bulge. 

\begin{figure} [ht]
\begin{center}
\includegraphics[width=\columnwidth]{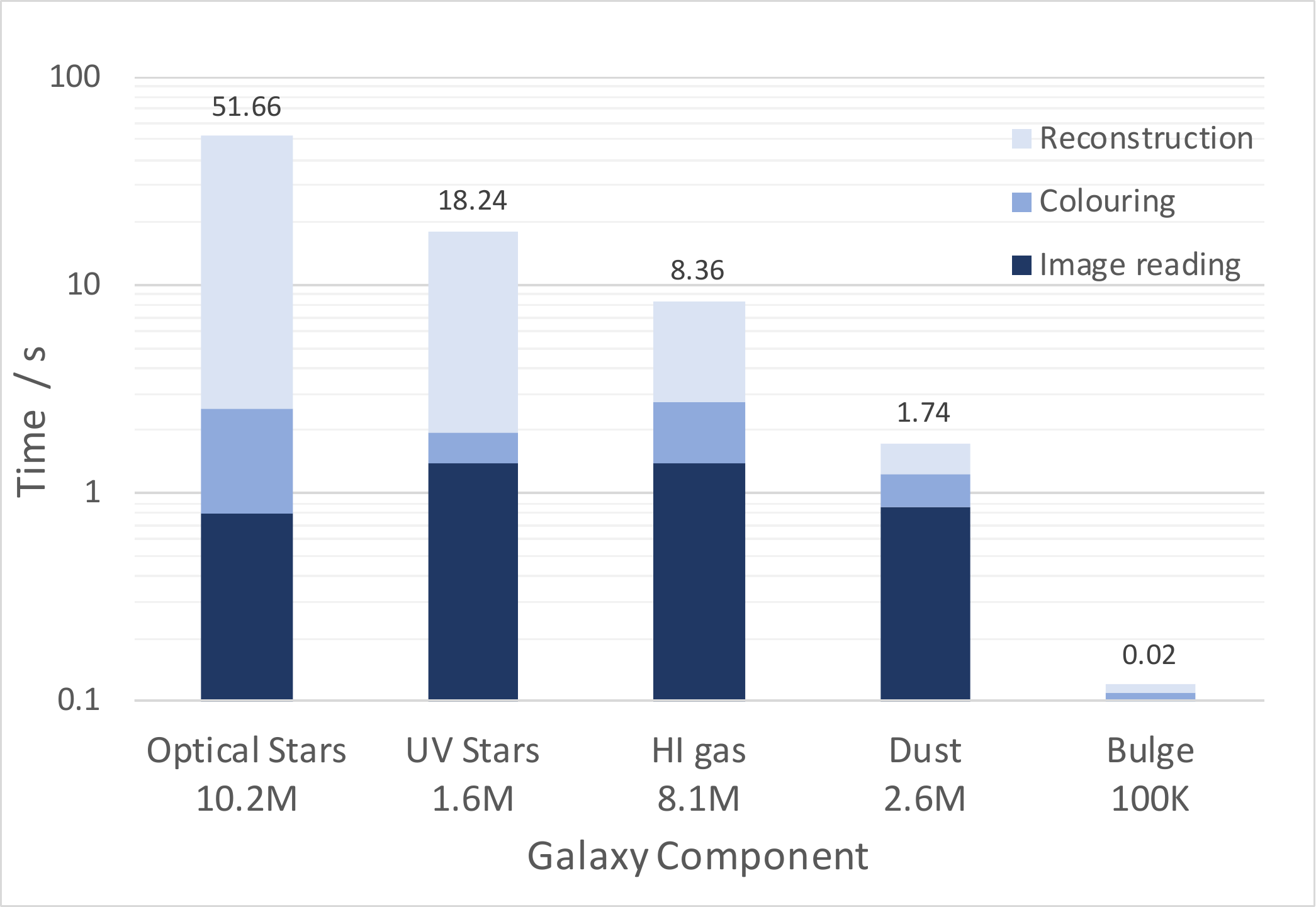} 
\caption{Performance of the galaxy reconstruction code for the M\,83 example case. The number of particles for each component is listed under the horizontal axis component labels.}
\label{fig:reconstruction-timing}
\end{center}
\end{figure}

The reconstruction and visualisation was performed using a single dual-socket node of a Cray XC50, with two 22-core Intel Broadwell processors clocked at 2.2 Ghz, and 128 GB of DDR4-2400 memory. The generated M\,83 model consists of approx. 22 million particles, split amongst the components as shown in Figure \ref{fig:reconstruction-timing}, which depicts the per-component computational time required for reading source image files, reconstruction, and colouring, totalling $\approx90$ seconds including output file I/O. The reconstruction is performed only once, with each visualisation image generated using the resultant 3D model. As illustrated, the image reading phase is dependent on the number of images required (three for H\,{\sc i}, two for UV, for example). Reconstruction is not directly dependent on the number of particles used for the component, as the particle count for the constructed component is a result of the saturation of the input image and the parameters determining seed particle count and surrounding point clouds, as discussed in Section \ref{sec:modelling}.

\begin{figure} [ht]
\begin{center}
\includegraphics[width=\columnwidth]{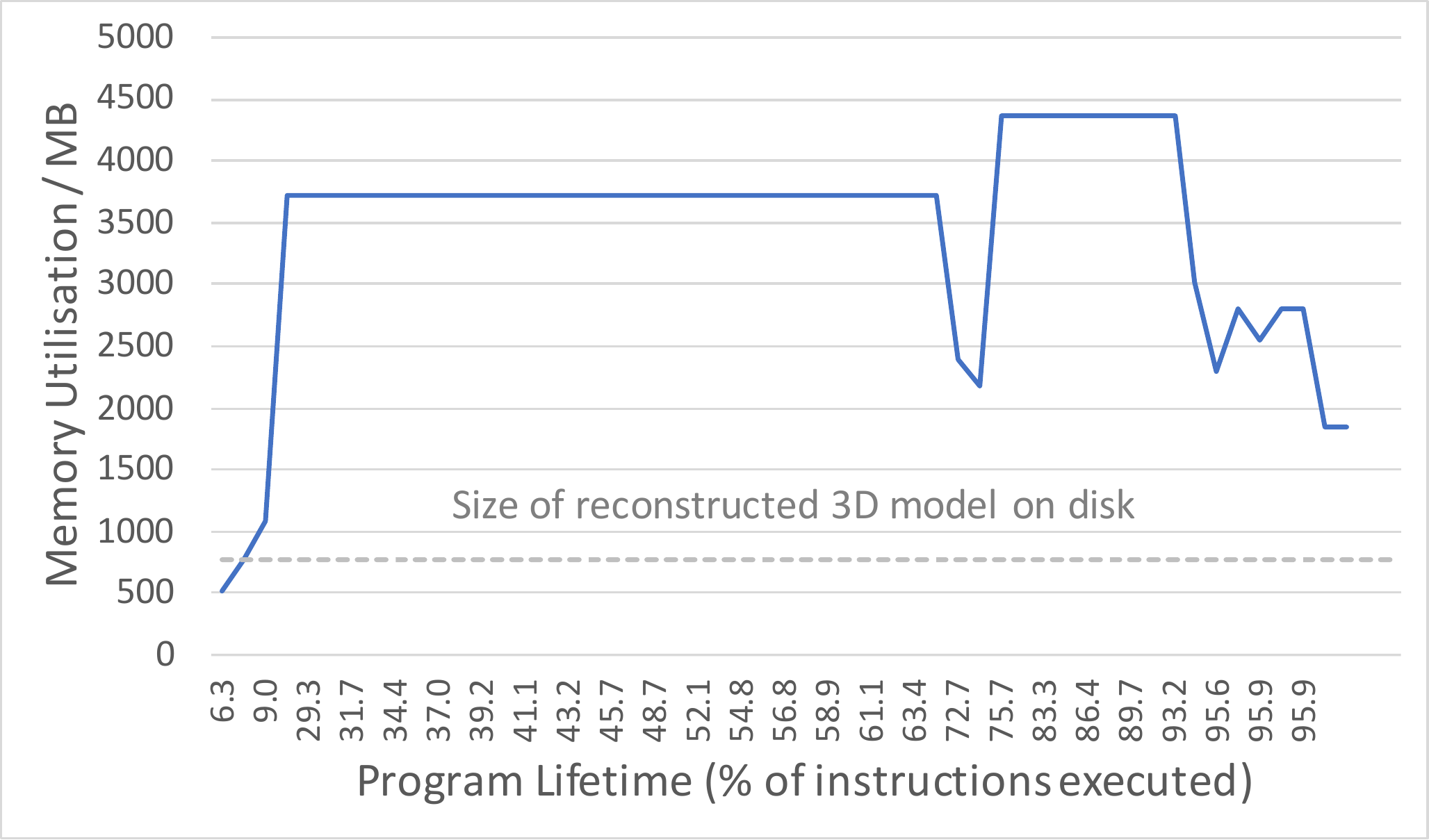} 
\caption{Memory consumption during program lifetime for the galaxy reconstruction code for the M\,83 example case.}
\label{fig:reconstruction-memory}
\end{center}
\end{figure}

The overall cost in terms of computational time for a single galaxy is, in general, acceptable; however, if this methodology were extended to model a large combined group with, for example, 10 to 1000 galaxies, the cost is expected to increase linearly, growing unreasonably high for 1000 galaxies. In this case, it is expected that non-interacting galaxies could be modelled independently to allow trivial parallelisation on a per-galaxy basis (e.g. one galaxy per-node of a compute cluster), whilst galaxy components could also be constructed in parallel to improve the per-galaxy computational cost (e.g. one or more cores of a compute node per galaxy component). Similarly, Figure \ref{fig:reconstruction-memory} shows that memory consumption can extend to approximately 5-6 times the size of the generated model throughout the program lifetime, which is acceptable for a single galaxy but a large group of combined galaxies processed in parallel may require further efforts to reduce the memory footprint. The plateaus seen in Figure \ref{fig:reconstruction-memory} are caused by allocating a large block of memory for particle data, and reusing this block throughout the reconstruction process for different components. 

\begin{figure} [ht]
\begin{center}
\includegraphics[width=\columnwidth]{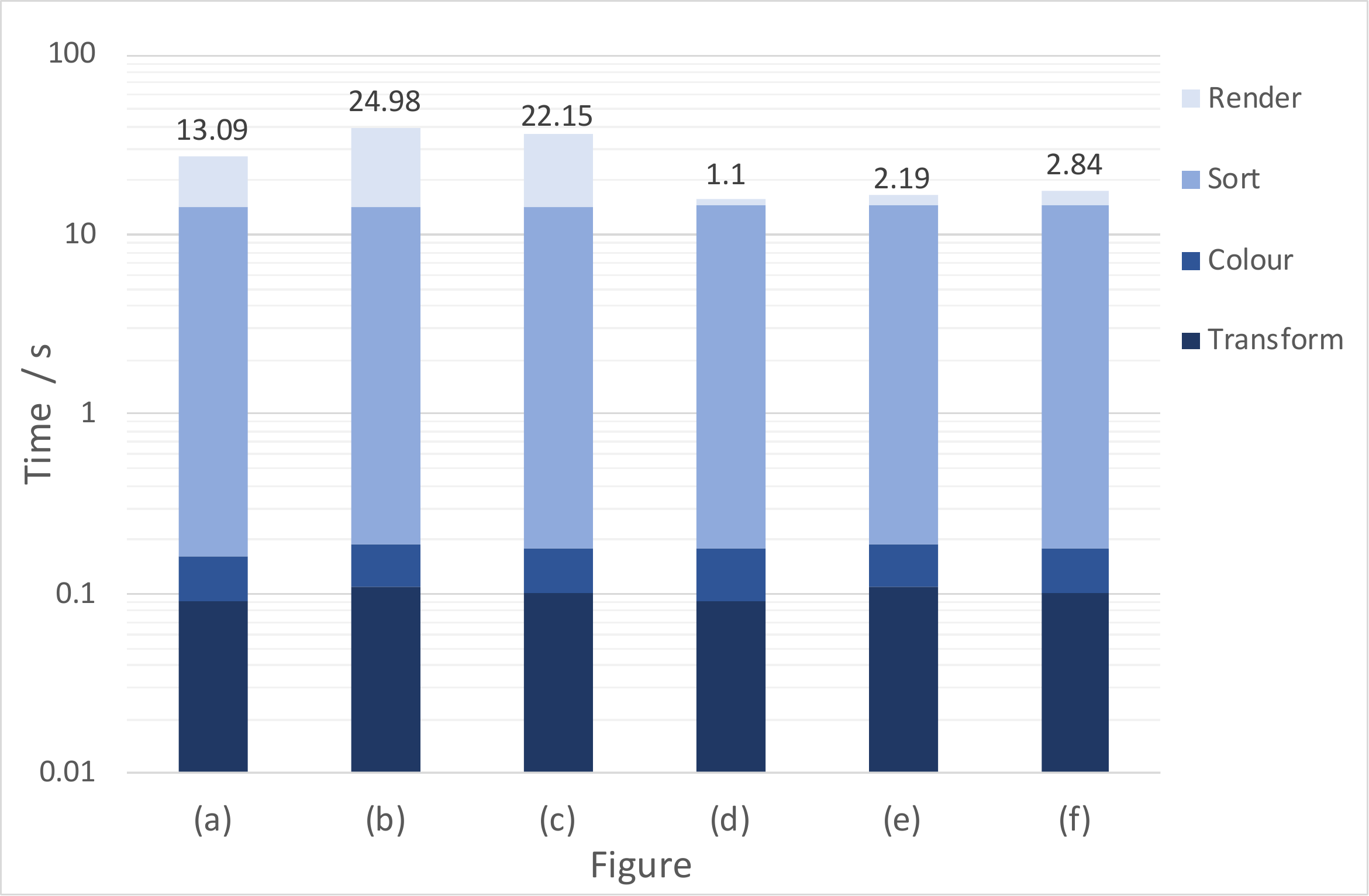} 
\caption{Rendering performance of the Splotch code for the M\,83 galaxy images seen in subfigures (a,b,c,d,e,f) of Figure \ref{fig:M83_Visualisation_results}. The galaxy model consists of approximately 22 million particles.}
\label{fig:visualisation-timing}
\end{center}
\end{figure}

Visualisation is performed using the Splotch software built with OpenMP support, and run with 44 OpenMP threads to match the number of cores available on the test platform. Figure \ref{fig:visualisation-timing} shows the visualisation time for each of the sub-figures of figure \ref{fig:M83_Visualisation_results}, split into the key components of the visualisation process. Transformation and colouring of particles are highly parallel operations taking $\approx 0.1s$ independent of the scene. Sorting is more computationally intensive, requiring $\approx 15s$ per frame, also inherently independent of the scene configuration. Rendering is scene-dependent, ranging from 1-3 seconds per scene for the stellar disk to 10-25 seconds per scene when the gaseous H\,{\sc i} disk is visible. This scene dependence stems from the average radius of particles, as discussed in \cite{RiviEtAl14}\cite{Dykesetal17}, a larger particle radius affects more pixels in the output image, which reduces parallel rendering performance. From these results, we can expect that during an extended movie rendering the gaseous H\,{\sc i} scenes will take approximately 10x the time of stellar scenes, as larger particles are utilised to create the gaseous effect seen in the H\,{\sc i} scene.

\begin{figure*}
\centering
\begin{tabular}{cc}
\includegraphics[width=0.8\columnwidth]{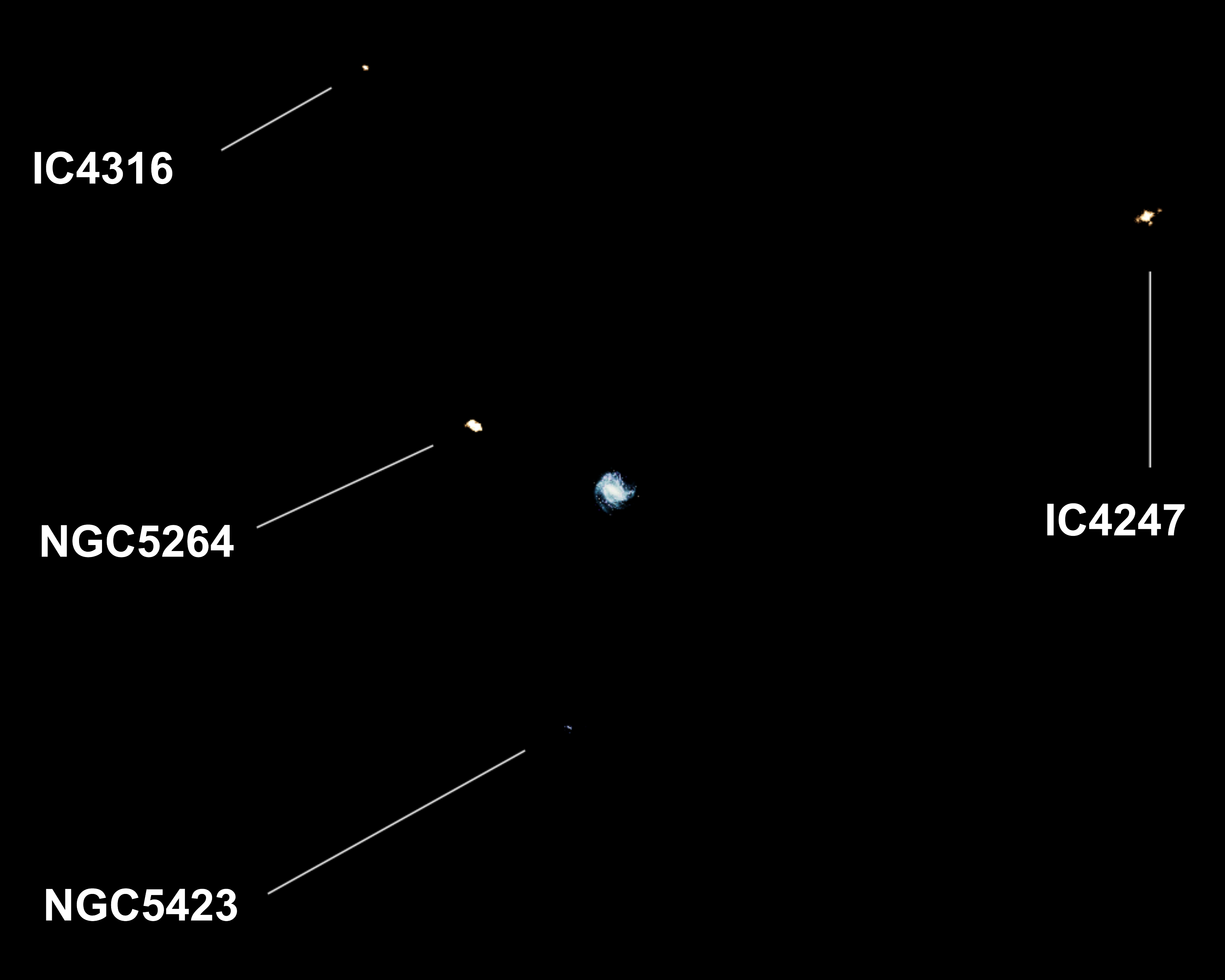} &
\includegraphics[width=0.8\columnwidth]{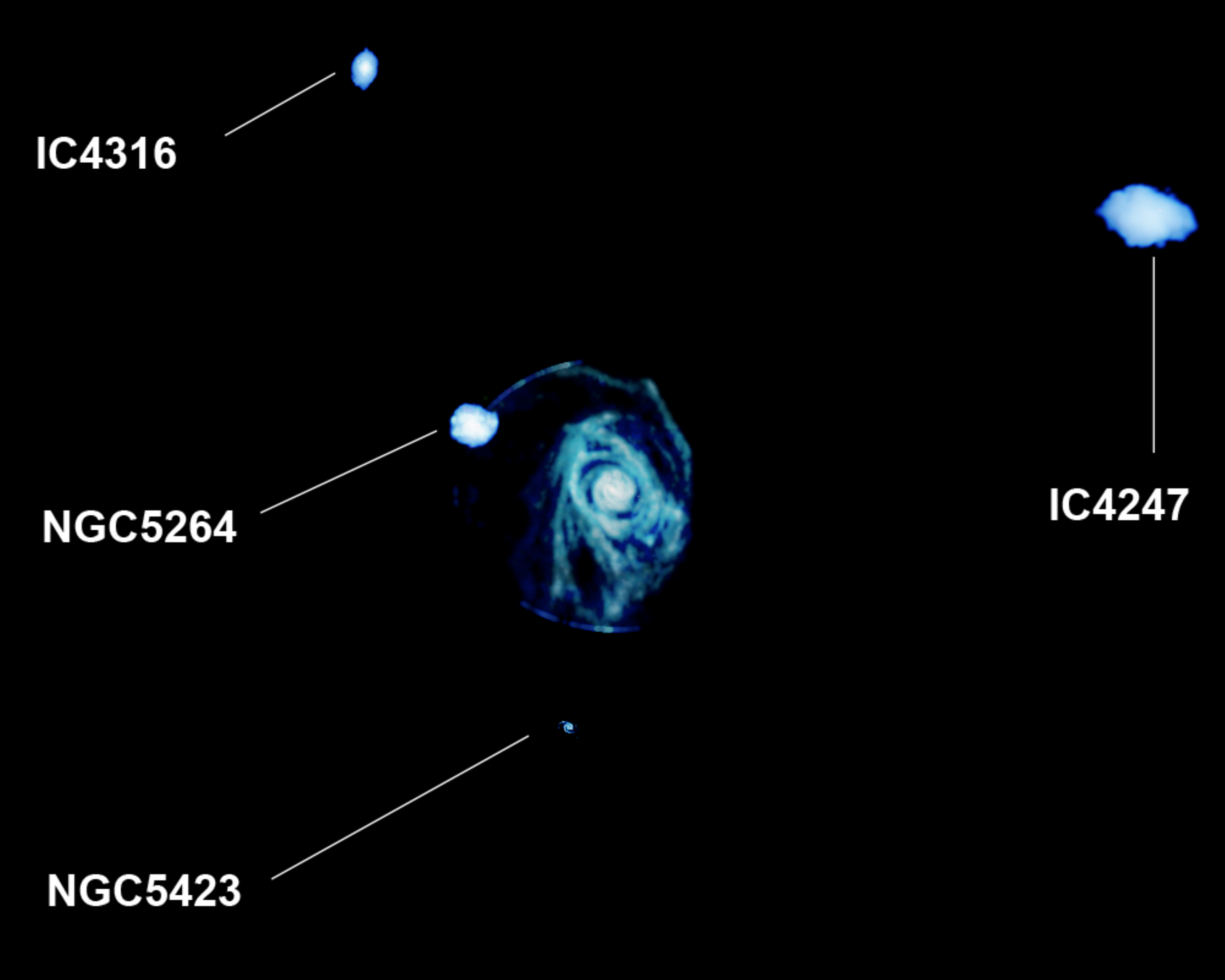} \\
\includegraphics[width=0.8\columnwidth]{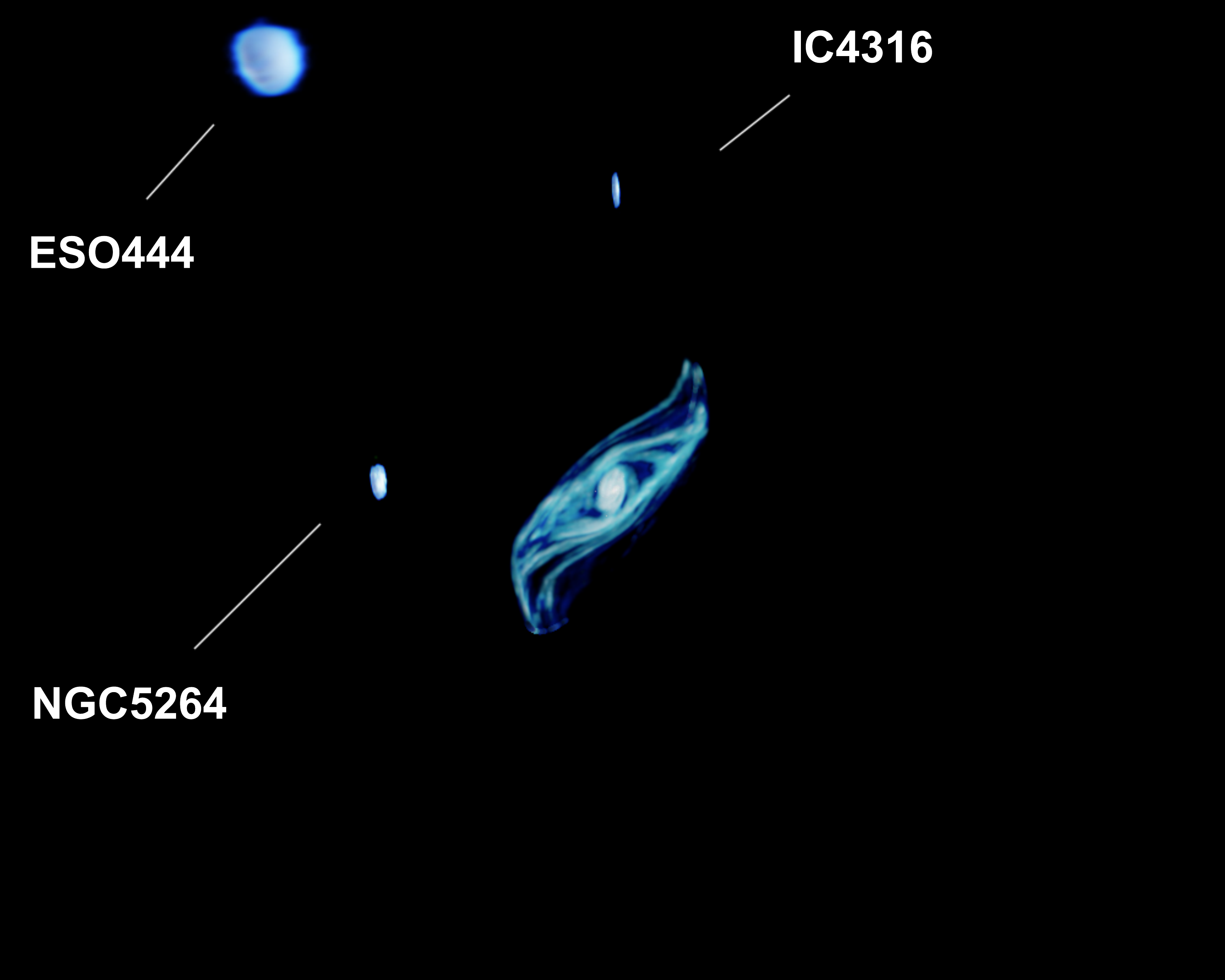} &
\includegraphics[width=0.8\columnwidth]{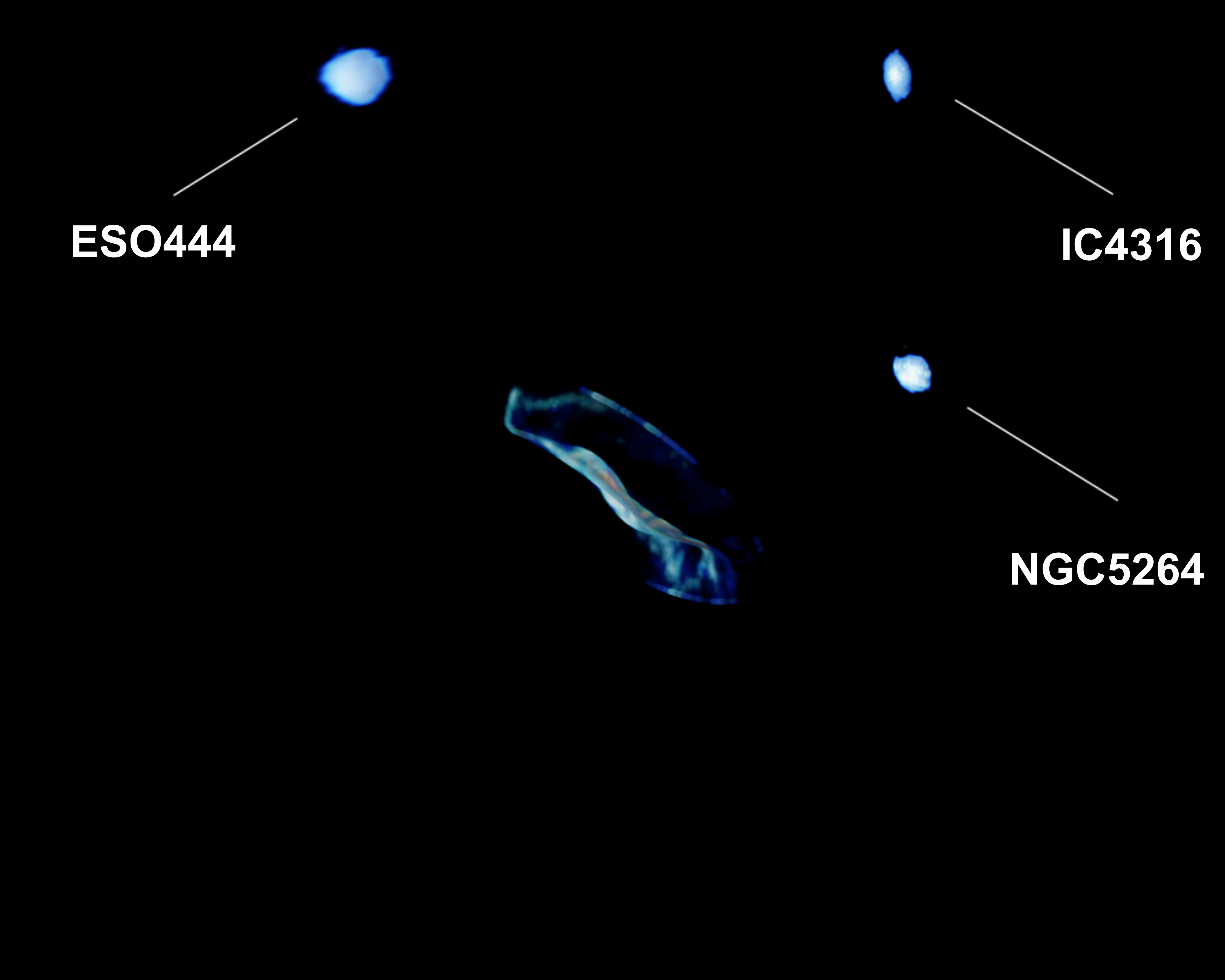} \\
\includegraphics[width=0.8\columnwidth]{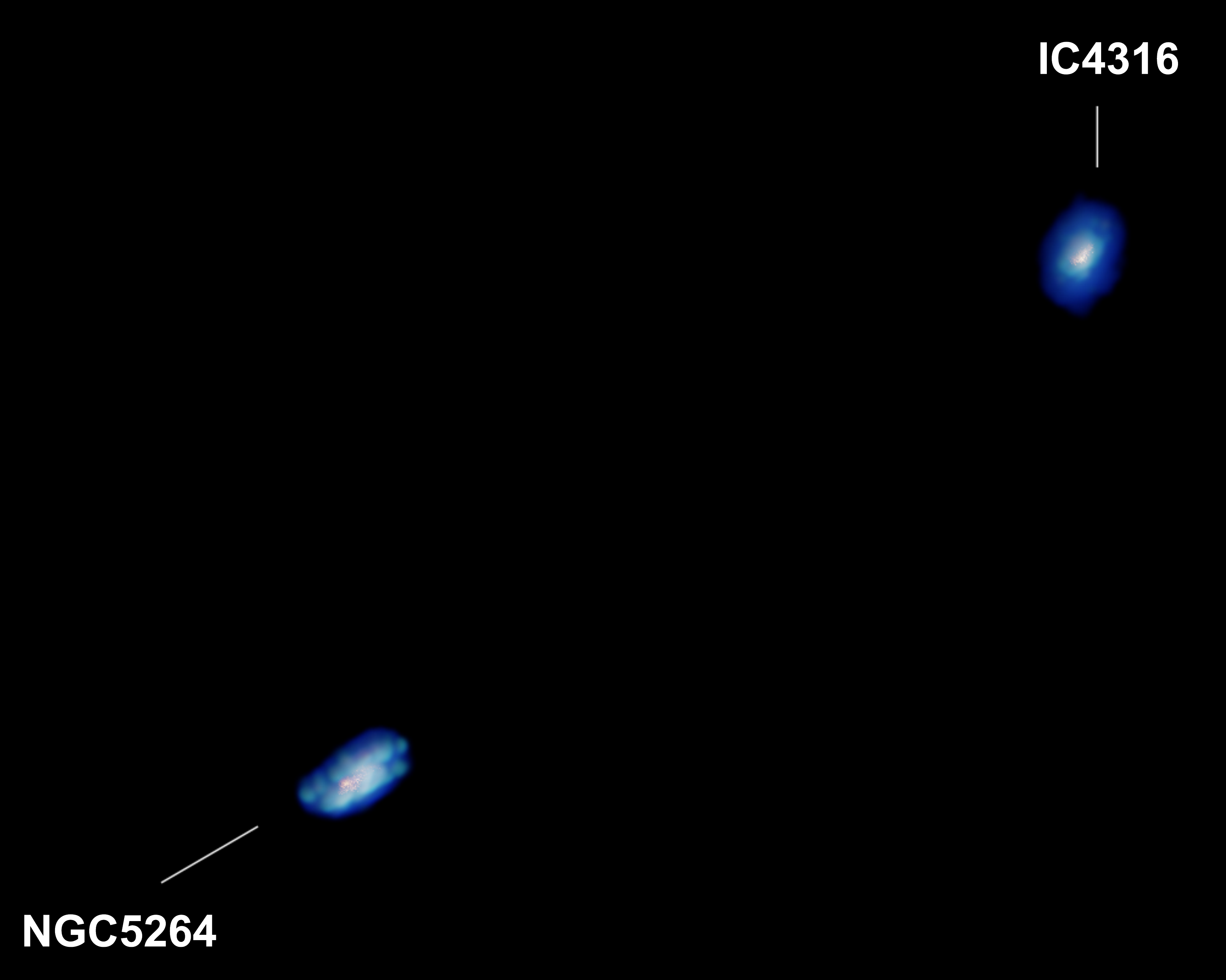} &
\includegraphics[width=0.8\columnwidth]{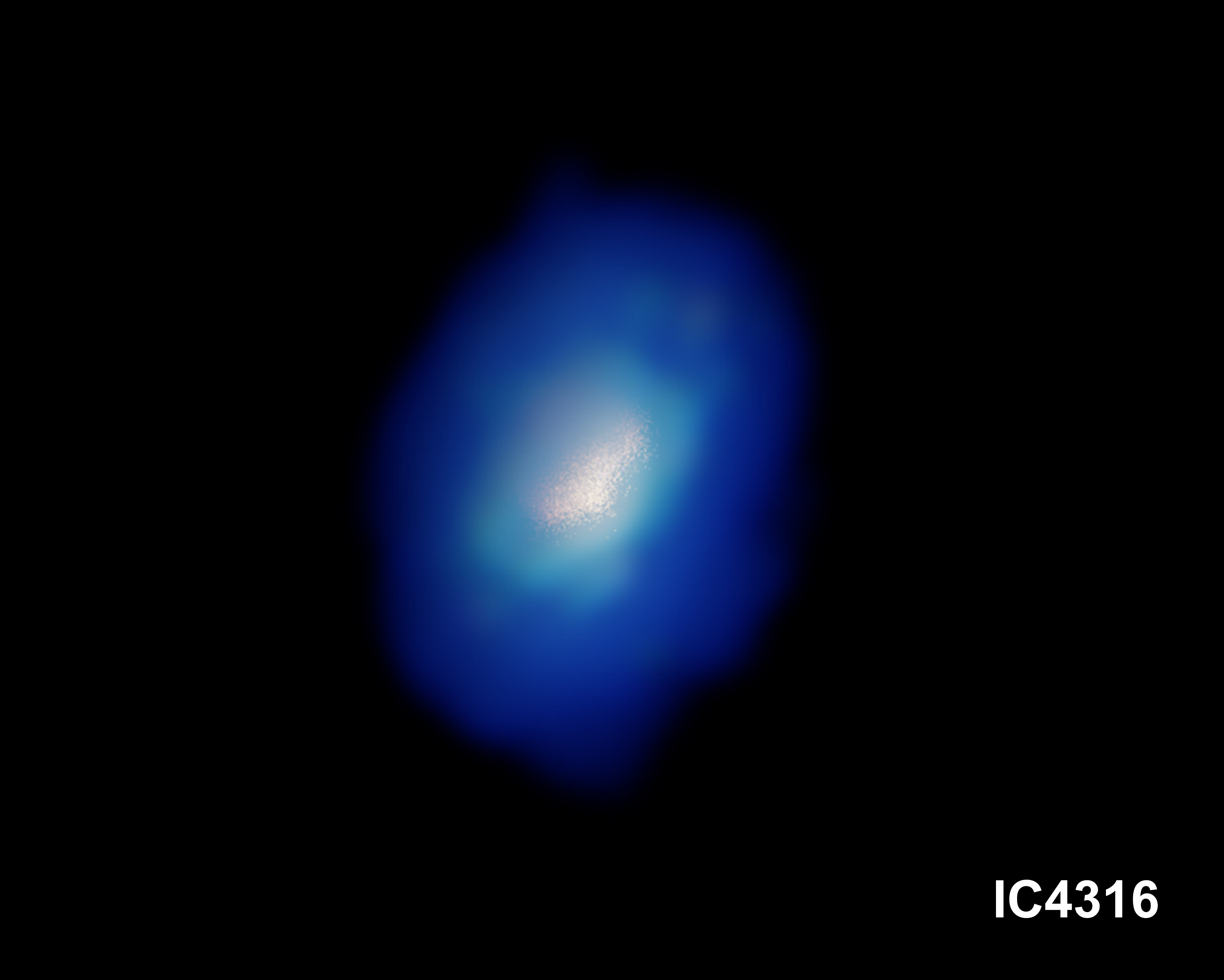} \\
\end{tabular}
\caption{The M\,83 galaxy and its neighbouring dwarf galaxies. Top left, the M\,83 group in the optical band. Top right, the M\,83 group with H\,{\sc i} emission added. Middle panels, the M\,83 group from different observation angles. Bottom left, two of the dwarfs companions of M\,83. Bottom right, a close up of dwarf companion IC4316. As this is a 3D rendering with multiple objects at different distances, we do not include a linear scaling in these figures.}
\label{fig:group}
\end{figure*}

Figure \ref{fig:group} demonstrates another view of M\,83, as a member of a group of galaxies known as the M\,83 local group. The top four panels show M\,83 as the most prominent object of group, alongside neighbouring dwarf galaxies, from a variety of viewing angles. According to observations, these galaxies may have either a disk-like shape or an almost spherical symmetry, and they have been reconstructed by our methodology following the procedure described in Section \ref{sec:PMPV_3DVis_Model} for dwarf galaxies. The source data for these objects is not high resolution, but sufficient to give an indication of their main features and distinguish between disk or elliptical galaxies for visualisation both in optical and in H\,{\sc i} (no data is available for possible dust distributions). The top-left panel shows the galaxy group from an Earth-based observing direction, centred on M\,83 (the brightest object), and including the star distribution of the four closest members of the local group. In the optical band the galaxies are just visible dots. However, as soon as H\,{\sc i} is added (top right panel), a much richer scene appears from the same point of view, with M\,83 showing the complex H\,{\sc i} distribution already highlighted above and the four dwarfs all clearly visible, with the gas distribution much more extended than the stellar distribution. The bottom left panel show a zoom-in of two of the members of the local group, a disk galaxy NGC5264 on the left, and IC4316 on the right. The bottom right panel shows a further zoom in of the dwarf galaxy IC4316, showing a more detailed view of the mixed star distribution and surrounding H\,{\sc i} cloud. 

The two middle panels of Figure \ref{fig:group} show the galaxy group from two different points of view, neither of which are Earth-based, demonstrating a unique view of these objects that is only possible with a reconstruction and visualisation methodology such as ours. These novel views highlight that the visual impression of galaxies shape, position, and structure, relative to one another can be misleading, due to projection effects of a 2D-only view. Furthermore, these different representations can provide hints to astronomers regarding features of the system, such as evolutionary interaction between neighbours. One possible explanation for the gas structure of M\,83 could be that past interactions with other galaxies in the local group cause the observed tilted and elongated tails which are not visible in the star distribution. Viewing the middle left panel could suggest that the upper gas tail of M\,83 may relate to a past interaction with the dwarf immediately above it. However, from the alternate point of view presented in the middle right panel, we observe that the tail is no longer pointing to the same dwarf, which moved to the top right corner of the image. Hence, a direct influence of the dwarf on M\,83's gas distribution appears unlikely, directing the scientist toward alternative explanations.

\section{Physical Realism}
\label{sec:physicalrealism}

\begin{table*}[]
\centering
\resizebox{\textwidth}{!}{%
\begin{tabular}{|l|l|l|l|l|l|}
\hline
\multicolumn{1}{|c|}{\textbf{\begin{tabular}[c]{@{}c@{}}Galaxy \\ Component\end{tabular}}} & \multicolumn{5}{c|}{\textbf{Component Property}} \\ 
\hline
\textbf{} & \multicolumn{1}{c|}{\textbf{\begin{tabular}[c]{@{}c@{}}Radial seed \\ distribution\end{tabular}}} & \multicolumn{1}{c|}{\textbf{\begin{tabular}[c]{@{}c@{}}Axial seed\\  distribution\end{tabular}}} & \multicolumn{1}{c|}{\textbf{\begin{tabular}[c]{@{}c@{}}Particle size \\ and distribution\end{tabular}}} & \multicolumn{1}{c|}{\textbf{Emission}} & \multicolumn{1}{c|}{\textbf{Absorption}} \\ 
\hline
Bulge & Observed model & Heuristic model & Heuristic model & Heuristic model & None \\ 
\hline
Stellar & Source data & Observed model & Heuristic model & \begin{tabular}[c]{@{}l@{}}Scaled to \\ source data\end{tabular} & None \\ 
\hline
Dust & Source data & Heuristic model & Heuristic model & \begin{tabular}[c]{@{}l@{}}Scaled to \\ source data\end{tabular} & \begin{tabular}[c]{@{}l@{}}Scaled to \\ source data\end{tabular} \\ 
\hline
Diffuse gas & Source data & Theoretical model & Heuristic model & \begin{tabular}[c]{@{}l@{}}Scaled to \\ source data\end{tabular} & None \\ 
\hline
\end{tabular}%
}
\caption[Summary of physical basis for galaxy model properties.]{A summary of the underlying models used for the various configurable properties of each galaxy component.}
\label{tab:ComponentProperties}
\end{table*}

\begin{figure*} [ht]
\centering
\includegraphics[width=0.8\textwidth]{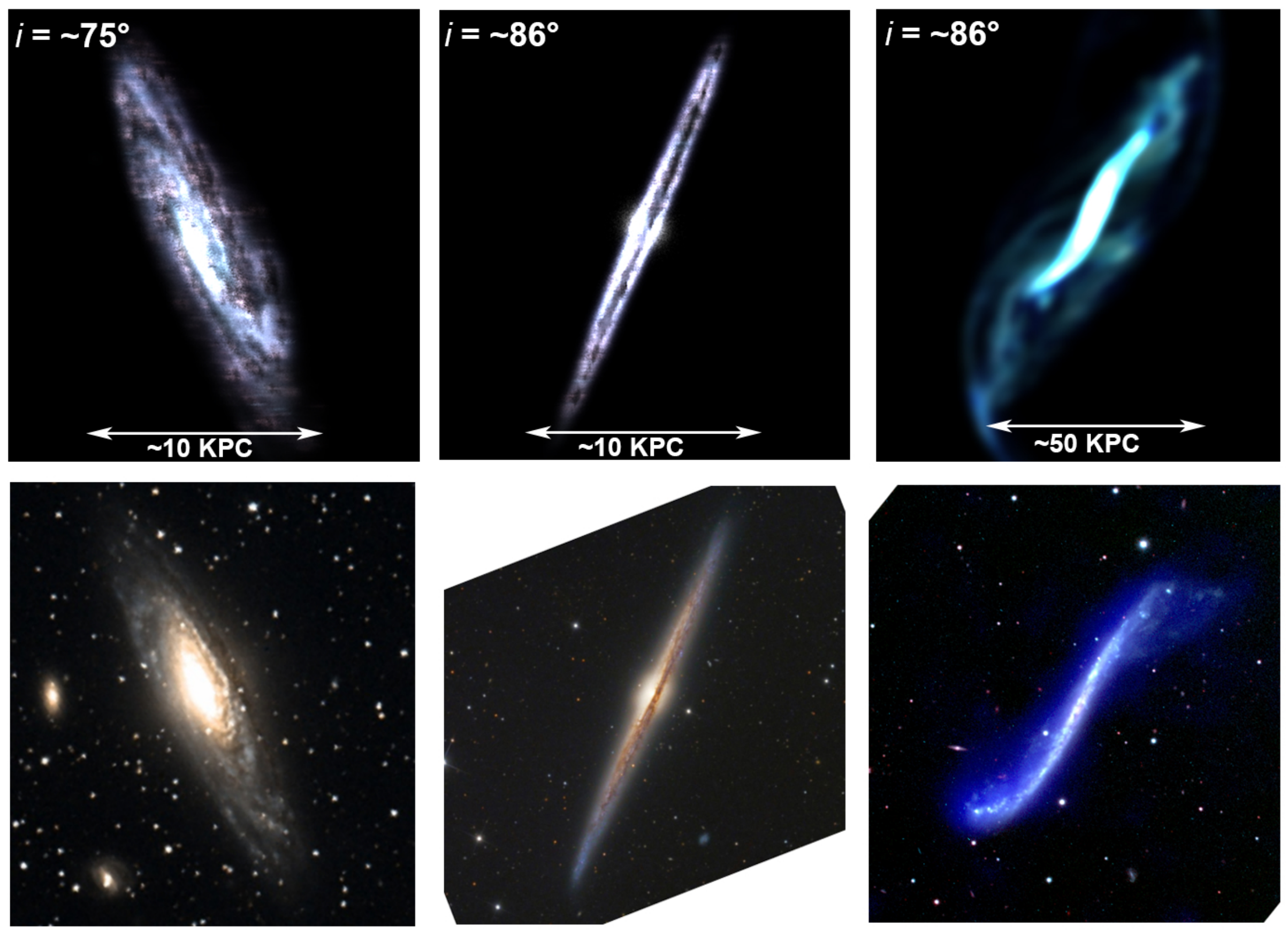}
\caption{A comparison of our M83 visualisations from non-Earth-based viewpoints to similar observations of other galaxies from Earth. The observed images (bottom row) have been rotated and scaled to match our images for comparison. On the left, we compare our visualisation of the M83 dusty stellar disk (\emph{top left}) to NGC7331, a two-color composite image (\emph{bottom left}) from the Digitised Sky Survey (see acknowledgements) extracted using Aladin Lite \cite{Bochetal14}. In the center, we compare our visualisation of the edge-on M83 stellar disk (\emph{top center}) to NGC~4565 (\emph{bottom center}), acknowledged as in Figure \ref{fig:M83_NGC4565}. On the right we compare our visualisation of the warped M83 H\,{\sc i} gas disk (\emph{top right}) with the warped hydrogen disk of UGC 3697 (\emph{bottom right}), courtesy of NRAO/AUI/NSF. }
\label{fig:ConstructedObservedComparison}
\end{figure*}

Our presented methodology aims to improve the analysis of the structure of galaxies based on combining kinematic and image information for observed galaxies. In general, it is not feasible to observe galaxies from an angle other than that of an Earth-based observer, and as such impossible to be certain that our model is structurally accurate for the observed galaxy. One approach for validation is to compare to observations of galaxies seen from Earth at similar viewpoints. Figure \ref{fig:ConstructedObservedComparison} compares a sample of our generated M83 images from non-Earth based viewpoints to similar shaped galaxy views extracted from observational data, highlighting the capability of our methodology to achieve a realistic result based on similar observations of other galaxies. 
  
We aim to build this physical realism into the galaxy representation using state of the art multi-wavelength observational data complemented by  numerical representations of analytical models for different components of galaxies as inferred from observations. To evaluate the extent to which this is achieved during construction, a series of high level morphological and visual properties have been identified with an aim to comprehensively describe the reconstruction process. The methods used to determine these properties for each galaxy component have been categorised by the extent to which physically realistic methods are utilised. The properties, with reference to the previous sections, are:

\begin{description}
\item[Radial seed distribution] This property defines the distribution of seed particles (as discussed in Section \ref{sec:PMPV_3DVis_Model}) in the radial plane of the galaxy (i.e. distribution across the galaxy face).
\item[Axial seed distribution] This property defines the distribution of seed particles in the axial plane of the galaxy (i.e. galaxy thickness).
\item[Particle size and distribution] This property defines the size of each representative particle and distribution of such particles around the seed (i.e. galaxy density).
\item[Emission] This property defines the value used as the emission coefficient (as discussed in Section \ref{sec:splotch}) for each particle during visualisation (i.e. galaxy brightness and colour).
\item[Absorption] This property defines the value used as the absorption coefficient for each particle during visualisation (ie. galaxy absorption, dust lanes, and shadows).
\end{description}

The method or model used to define each of these properties per galaxy component has been categorised into one of the following groups, ordered by physically realism, with the top being most realistic:

\begin{description}
\item[Source data] The property has been derived directly from observational data of the source object being modelled.
\item[Observed model] The property has been derived from general observations or relations observed for the object, or type of object, being modelled.
\item[Theoretical model] The property has been derived from a theoretical model or simulation of the object, or type of object, being modelled.
\item[Scaled to source data] The property has been scaled according to source data, but may be initially defined by heuristic. 
\item[Heuristic model] The property has been defined and tuned heuristically in collaboration with an experienced astronomer based on resultant visual effects.
\end{description}

As illustrated in Table \ref{tab:ComponentProperties}, whilst much of the modelling and visualisation methodology exploits observed or inferred data, there are a series of heuristics requiring manual intervention which introduces a bias in our results. In particular, the initial number of particles per component is currently empirically tuned based on visualisation results, we envision a more robust approach to determining initial particle counts considering observed parameters (e.g. estimated mass of each component) for the galaxy being modelled. Further work is envisioned to develop a more realistic axial dust model, considering the existing research into the morphology and scattering effects of dust grains in both in astronomy and visualisation research such as \cite{Magnoretal05}, who use a 3D Perlin noise model \cite{Perlin85} to describe dust morphology combined with careful treatment of dust scattering, or \cite{Nadeau01} who similarly use Perlin noise for a more realistic gaseous nebulae effect. Furthermore, the axial distribution of the bulge component may be more robustly defined based on observed properties in systematic studies of bulges in galaxies, for example \cite{Gadotti09, Fisher11}.

\section{Conclusions and Future Work}
\label{sec:conclusion}


We have presented a novel 3D modelling and visualisation methodology for the reconstruction of nearby galaxies based on a wide variety of multi-resolution observed images and derived data. This includes optical, UV, IR and H\,{\sc i} images along with H\,{\sc i} kinematic models of nearby galaxies based on tilted-ring fitting. We addressed several objectives: (1) to create realistic views of galaxies from viewpoints that are not otherwise possible to observe from, (2) explore the validity of derived spatial 3D models of such galaxies, and (3) allow enhanced visual analysis of the 3D galaxy morphology (stars, gas and dust). Our methodology allows dynamic 3D exploration of galaxies and can provide new views (see Figure \ref{fig:M83_Visualisation_results}) of galaxies that are only typically observable from an Earth-based viewpoint. Such novel visualisations are able to illustrate several different galaxy components, e.g., extended H\,{\sc i} disks and inner stellar disks, only the latter of which is typically well known to the public, and show the 3D structure of the extended H\,{\sc i} disk of M83, incorporating kinematic models in our reconstruction to represent the warped structure that is not typically visible in 2D observed images. 

We further enable comparison of 3D reconstructed galaxies with other observed galaxies at different viewing angles (as in Figure \ref{fig:ConstructedObservedComparison}), which could support evaluation of the kinematical models that inform their 3D shapes. Our methodology allows enhanced analysis of these galaxies, e.g. allowing the viewer to distinguish whether observed features are real characteristics or projection effects, or highlight properties hidden by a 2D representation, as demonstrated in Figure \ref{fig:group}. We note that galaxies observed close to edge-on can be difficult to reconstruct in 3D as detailed information of their spiral structure is not easily available, as such the current methodology is not able to present a realistic face-on view of observed edge-on galaxies. One approach to address this problem from a visual perspective could be to generate a believable face-on impression through statistical analysis of existing observations of face-on galaxies, or exploiting recent results of high resolution cosmological simulations such as EAGLE \cite{Mcalpineetal16} to realistically describe the spiral arms in galaxies. 

Work is on-going focusing on introducing more automated pipeline stages, enabling new galaxies to be modelled faster and reducing the bias introduced by manual intervention as discussed in Section \ref{sec:physicalrealism}. The existing pipeline for both the modelling and visualisation is command-line based, with key-value parameter files as inputs; to improve uptake and streamline the modelling and visualisation process we are considering the viability of a more user-friendly interface such as a Python scripting layer or graphical interface. We then intend to release the modelling code, alongside full instructions for source data collection and preparation, alongside the next major release of the Splotch visualisation code\footnote{\url{https://github.com/splotchviz/splotch}}, for use on Linux and MacOS systems.

In future, we plan to enable interactive visualisations and fly-throughs of the 3D reconstructed galaxies and galaxy groups. Such fly-through movies, or even interactive exploration of single galaxies or galaxy groups, are envisaged to provide new opportunities for visual discovery such as those discussed in Section \ref{sec:results}. We believe it may also be interesting to explore the combination of our modelling approach with more cinematic rendering tools, such as utilised in the works of \cite{Naiman17,Borkiewicz19}, relaxing the requirements on physical realism to improve visual impact for outreach purposes. 

Finally, our galaxy group demonstration is the first step toward a comprehensive methodology able to provide a visual representation of large galaxy assemblies based on their observational properties. A survey-like approach will require streamlining of the manual, user-defined, stages of our pipeline, which would quickly become the most limiting scaling factor, however holds significant analysis opportunities and could, for example, lead to a generally accessible dataset holding many 3D referential images of a variety of galaxies. We further hope that in future this may lead to more advanced analysis scenarios of dynamic scenes, for example exploring a dynamic scene of moving galaxy neighbours that demonstrates possible past or future states of a group based on observed inter-galaxy kinematics.

\section*{Acknowledgements}

KD acknowledges support by the ORIGINS cluster, funded by the Deutsche Forschungsgemeinschaft (DFG, German Research Foundation) under Germany's Excellence Strategy, EXC-2094, 390783311. MK acknowledges support by NEANIAS, funded by the EC Horizon 2020 research and innovation programme under grant agreement No. 863448.

Figure \ref{fig:ConstructedObservedComparison} is based on photographic data of the National Geographic Society Palomar Observatory Sky Survey (NGS-POSS) obtained using the Oschin Telescope on Palomar Mountain. The NGS-POSS was funded by a grant from the National Geographic Society to the California Institute of Technology. The plates were processed into the present compressed digital form with their permission. The Digitized Sky Survey was produced at the Space Telescope Science Institute under US Government grant NAG W-2166.

\bibliographystyle{model2-names}
\bibliography{biblio}







\end{document}